\documentclass[pdflatex,sn-mathphys-num]{sn-jnl}
\usepackage{graphicx}%
\usepackage{subcaption}
\usepackage{float}
\usepackage{placeins}
\usepackage{multirow}%
\usepackage{amsmath,amssymb,amsfonts}%
\usepackage{amsthm}%
\usepackage{mathrsfs}%
\usepackage[title]{appendix}%
\usepackage{xcolor}%
\usepackage{textcomp}%
\usepackage{manyfoot}%
\usepackage{booktabs}%
\usepackage[utf8]{inputenc}
\usepackage{textgreek}
\usepackage{algorithm}%
\usepackage{algorithmicx}%
\usepackage{algpseudocode}%
\usepackage{listings}%
\usepackage{tabularx}%
\theoremstyle{thmstyleone}%
\theoremstyle{thmstyletwo}%

\theoremstyle{thmstylethree}%

\begin{document}

\title[Article Title]{A Dual-Memory Ferroelectric Transistor Emulating Synaptic Metaplasticity for High-Speed Reservoir Computing}


\author[1]{\fnm{Yifan} \sur{Wang}}\email{wang4838@purdue.edu}

\author[2]{\fnm{Muhammad Sakib} \sur{Shahriar}}\email{mshahria@go.olemiss.edu}

\author[3]{\fnm{Salma} \sur {Soliman}}\email{ssoliman7@gatech.edu}

\author[4]{\fnm{Noah} \sur{Vaillancourt}}\email{nvaillan@purdue.edu}

\author[3]{\fnm{Lance} \sur{Fernandes}}\email{lfernandes33@gatech.edu}

\author[5]{\fnm{Andrea} \sur{Padovani}}\email{andrea.padovani@unimore.it}

%\author[6]{\fnm{Valerio} \sur{Lunardelli}}\email{Valerio\_Lunardelli@amat.com}

%\author[6]{\fnm{Luca} \sur{Larcher}}\email{Luca\_Larcher@amat.com}

%\author[7]{\fnm{Gaurav} \sur{Thareja}}\email{Gaurav\_Thareja@amat.com}

\author[3]{\fnm{Asif Islam} \sur{Khan}}\email{akhan40@gatech.edu}

\author*[2]{\fnm{Md Sakib} \sur{Hasan}}\email{mhasan5@olemiss.edu}

\author*[4,1]{\fnm{Raisul} \sur{Islam}}\email{raisul@purdue.edu}

\affil[1]{\orgdiv{Elmore Family School of Electrical and Computer Engineering}, \orgname{Purdue University}, \orgaddress{\city{West Lafayette}, \state{IN}, \country{USA}}}

\affil[2]{\orgdiv{Department of Electrical and Computer Engineering}, \orgname{University of Mississippi}, \orgaddress{\city{University}, \state{MS}, \country{USA}}}

\affil[3]{\orgdiv{School of Electrical and Computer Engineering}, \orgname{Georgia Institute of Technology}, \orgaddress{\city{Atlanta}, \state{GA}, \country{USA}}}

\affil[4]{\orgdiv{School of Materials Engineering}, \orgname{Purdue University}, \orgaddress{\city{West Lafayette}, \state{IN}, \country{USA}}}

\affil[5]{\orgdiv{Dipartimento di Scienze e Metodi dell'Ingegneria}, \orgname{Università di Modena e Reggio Emilia}, \orgaddress{\city{Reggio Emilia}, \state{RE}, \country{Italy}}}

%\affil[6]{\orgname{Applied Materials}, \orgaddress{\city{Reggio Emilia}, \state{RE}, \country{Italy}}}

%\affil[7]{\orgname{Applied Materials}, \orgaddress{\city{Sunnyvale}, \state{CA}, \country{USA}}}

\newpage
%%Compact 200 word abstract for AdvMat
\abstract{The exponential growth of edge artificial intelligence demands material-focused solutions to overcome energy consumption and latency limitations when processing real-time temporal data. Physical reservoir computing (PRC) offers an energy-efficient paradigm but faces challenges due to limited device scalability and reconfigurability. Additionally, reservoir and readout layers require memory of different timescales, short-term and long-term respectively—a material challenge hindering CMOS-compatible implementations. This work demonstrates a CMOS-compatible ferroelectric transistor using hafnium-zirconium-oxide (HZO) and silicon, enabling dual-memory operation. This system exhibits non-volatile long-term memory (LTM) from ferroelectric HZO polarization and volatile short-term memory (STM) from engineered non-quasi-static (NQS) channel-charge relaxation driven by gate–source/drain overlap capacitance. Ferroelectric polarization acts as non-volatile programming of volatile dynamics: by modulating threshold voltage, the ferroelectric state deterministically switches the NQS time constant and computational behavior between paired-pulse facilitation (PPF) and depression (PPD). This establishes a generalizable material-design principle applicable to diverse ferroelectric-semiconductor heterostructures, extending beyond silicon to oxide semiconductors and heterogeneously-integrated systems. The device solves second-order nonlinear tasks with 3.69 $\times$ 10$^{-3}$ normalized error using only 16 reservoir states—$\sim$5$\times$ reduction—achieving 20 $\mu$s response time ($\sim$1000$\times$ faster) and 1.5 $\times$ 10$^{-7}$ J energy consumption, providing an immediately manufacturable pathway for neuromorphic hardware and energy-efficient edge intelligence.}




\maketitle

\section{Introduction}\label{sec1}

Reservoir computing (RC) is a powerful and efficient computational paradigm for processing complex temporal and sequential data, offering a substantial reduction in training overhead compared to conventional recurrent neural networks (RNNs). Since the reservoir, functioning as the core nonlinear dynamic system, remains fixed and only the readout layer is trained, RC achieves much faster learning, resulting in a lower computational cost \cite{tanaka_recent_2019}. This streamlined training process, combined with the reservoir’s inherent ability to project inputs into a high-dimensional feature space, makes RC exceptionally well-suited for resource-constrained and low-latency edge-computing applications, including speech recognition \cite{takagi_physical_2023}, image classification \cite{song_reservoir_2023}\cite{nowshin_merrc_2024}, and temporal sensor-signals analysis like electrocardiogram (ECG)\cite{liedji_delay-based_2023}\cite{gaurav_spiking_2022}. 
\begin{figure}[tbhp]
    \includegraphics[width=\linewidth]{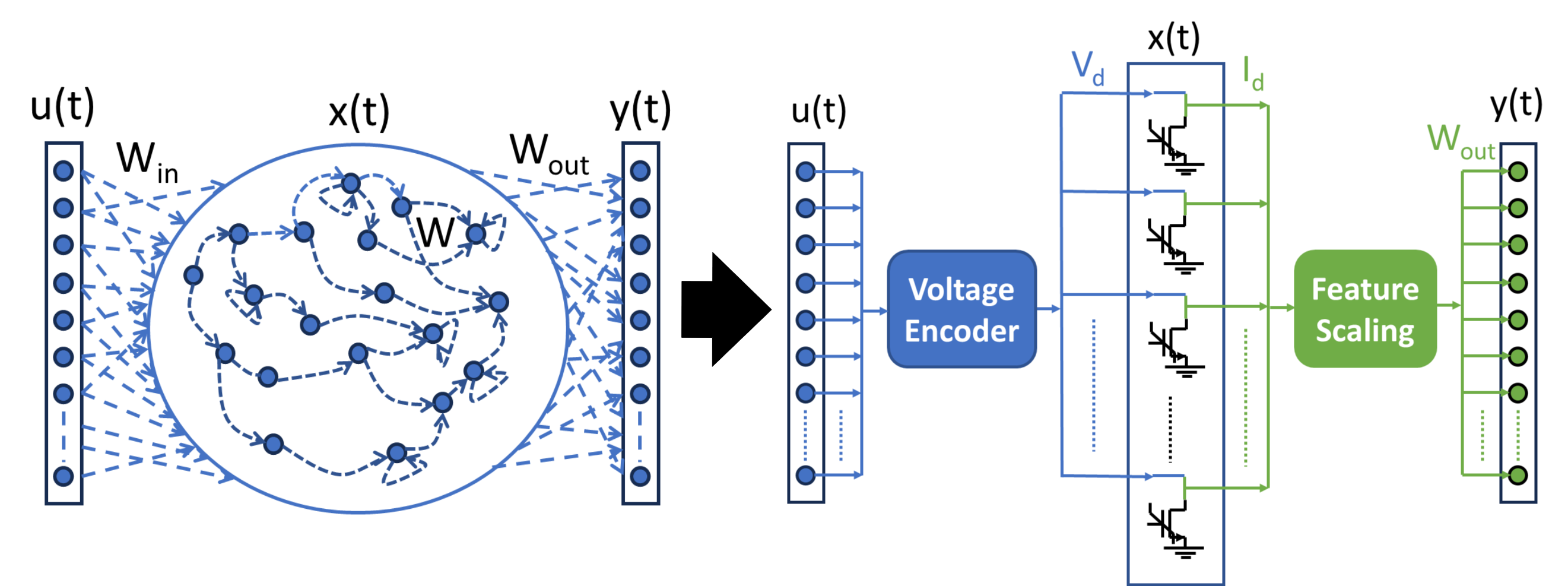}
    \caption{A representation of a conventional reservoir and how it is converted into the FeFET reservoir. In the conventional approach (left), the reservoir is modeled as a recurrent neural network; the input weights ($W_{in}$) and the recurrent reservoir weights ($W$) remain fixed, while only the readout weights ($W_{out}$) are trained. In contrast, the FeFET reservoir (right) comprises heterogeneous FeFET devices exhibiting distinct behavior. The input sequence u(t) is first encoded into voltage signals, which drive the FeFET array. The resulting output currents are scaled according to device conductance to produce the reservoir states, which are then processed by the readout layer.}
    \label{fig:FeFET_Reservoir}
\end{figure}
Physical reservoir computing (PRC) has emerged as a promising alternative to traditional RNNs, utilizing diverse nonlinear dynamic physical platforms, including electronic \cite{m_c_soriano_delay-based_2015}, \cite{schrauwen_compact_2008}, \cite{n_soures_robustness_2017}, photonic \cite{van_der_sande_advances_2017}, \cite{brunner_tutorial_2018}, mechanical \cite{hauser_towards_2011}, \cite{caluwaerts_design_2014}, biological \cite{dranias_short-term_2013}, \cite{sussillo_generating_2009}, quantum \cite{fujii_harnessing_2017}, and spintronic \cite{torrejon_neuromorphic_2017} mechanisms, to achieve rapid information processing with minimal learning costs. The fundamental principle is that any nonlinear dynamic system can function as a reservoir if it satisfies four key criteria \cite{tanaka_recent_2019}: (i) high dimensionality, (ii) nonlinearity, (iii) fading (short-term) memory \cite{maass_fading_2004}, \cite{yildiz_re-visiting_2012}, and (iv) the separation property, which ensures that distinct inputs are mapped to separable states in the high-dimensional space. Consequently, practical physical reservoir computing demands material platforms that simultaneously satisfy competing requirements: CMOS manufacturability, non-linear dynamics with volatile memory, non-volatile memory, and reconfigurable behavior—constraints that are challenging to implement together. Therefore, state-of-the-art implementations of PRC face significant challenges, including device-to-device or cycle-to-cycle variations, ensuring stable and consistent readouts, limited reconfigurability, and overcoming inherent physical limitations on processing timescales \cite{liu_physical_2025}, \cite{ma_spiking_2025}. Solving all challenges in a CMOS-compatible heterogeneously-integrated system has not been demonstrated so far using PRC systems that utilize exotic materials for non-linear dynamics.

In recent years, there has been growing research interest in CMOS-compatible material platforms for emulating neural behavior \cite{pazos_synaptic_2025}. Ferroelectric field-effect transistors (FeFETs) employing hafnium zirconium oxide (e.g. Hf$_{0.5}$Zr$_{0.5}$O$_2$/HZO) as the gate dielectric have emerged as one of the promising physical reservoir-computing platforms due to their fast, scalable memory capability and CMOS compatibility. A notable example is the silicon‑channel FeFET incorporating HZO, which uses ferroelectric polarization dynamics and polarization–charge coupling to effectively demonstrate short-term memory and high‑dimensional nonlinear transformations, successfully solving nonlinear time-series prediction tasks via simple regression models \cite{toprasertpong_reservoir_2022}. However, a key limitation of this approach is its reliance on binary or discrete input signals for complex tasks like the second-order nonlinear dynamic system (SONDS) prediction, wherein the utilization of discrete input signals defeats the purpose of employing PRC in real-time temporal signal processing. This constraint arises because the STM mechanism of this work relies on the gate-driven switching of ferroelectric domains, a process more naturally suited for distinct and non-volatile state changes rather than the volatility in storing the previous states required for continuous signals. In effect, this approach does not show a true STM but an effective STM through limited sampling using a temporal mask function. Additional demonstrations of physical reservoir computing based on ferroelectric polarization switching also encounter notable constraints. Liu \emph{et al.} reported a PRC system employing a ferroelectric semiconductor (α-In$_2$Se$_3$) channel transistor, constructed by vertically stacking multiple α-In$_2$Se$_3$ layers \cite{liu_multilayer_2022}. While functional, this architecture lacks CMOS compatibility and scalability. Furthermore, it does not permit independent modulation of short-term and long-term memory dynamics and suffers from a slow temporal response ($\sim$100 ms). Similarly, Ju \emph{et al.} demonstrated a PRC implementation using a HfAlO-based memristor \cite{ju_reservoir_2024}, in which ferroelectric polarization switching modulates the tunnel barrier, with partial polarization switching utilized to emulate short-term memory. However, the resulting relaxation times ($\tau$)—ranging from milliseconds to seconds—are prohibitively long, preventing the complete erasure of the conductive state at extended inter-pulse intervals.

Our research specifically exploits a FeFET structure integrating silicon channels with Hf$_{0.5}$Zr$_{0.5}$O$_2$ (HZO) ferroelectric material. The device uniquely incorporates dual memory modalities: long-term memory (LTM) derived from ferroelectric gate polarization switching and short-term memory (STM) arising from a \emph{non-quasi-static (NQS)} channel-charge response that is \emph{engineered} by the gate–source/drain overlap capacitance. Under appropriate drain-voltage pulsing, the overlap provides a localized capacitive coupling at the drain edge; because the inversion channel behaves as a distributed, bias-dependent RC network where the perturbed channel charge relaxes with a finite time constant, $\tau$,  equivalent to the relaxation time of the STM. (Details on the NQS model of transistors and its corresponding equivalent RC circuit model are detailed in Supplementary Fig. 10 and the Supplementary Note that follows.) This yields microsecond-scale STM expressed as tunable paired-pulse facilitation/depression (PPF/PPD) in the \(I_d\)–\(V_{ds}\) transients. Because the STM mechanism is overlap-capacitance–driven NQS rather than ferroelectric switching or trapping, it is intrinsically CMOS-scalable: relaxation time, \(\tau \!\sim\! R_{\mathrm{ch}}C_{\mathrm{ov}}\) can be tuned by geometry/materials (to set \(C_{\mathrm{ov}}\)) and bias/length (which set \(R_{\mathrm{ch}}\)), enabling high speed at low energy without sacrificing compatibility. As ferroelectric polarization modulates the effective threshold voltage, it selects the operating regime (strong versus weak inversion) and thereby programmatically tunes the sign and strength of the NQS STM under a fixed \(V_{gs}\). This co-action of LTM configuring STM is analogous to brain synaptic function, where long-term plasticity modulates short-term dynamics. By integrating both STM and LTM within a single FeFET device, we also enable reservoir-layer heterogeneity that can be adjusted during operation, promising enhanced reconfigurability for a broad range of applications. Our FeFET reservoir processes continuous data streams; the resulting device-level transients constitute the reservoir’s internal state, from which trained readouts produce computational outputs.

As a test case for this combined LTM+STM operation, we specifically demonstrate the capacity of our silicon–HZO FeFET reservoir to solve second-order nonlinear dynamic tasks with high accuracy and efficiency. We perform systematic device and bias optimization, characterize memory dynamics and nonlinearity, and benchmark performance. The results highlight the potential of co-integrated NQS STM and ferroelectric LTM in FeFET-based reservoir-computing architectures for next-generation, low-power edge computing. By co-engineering ferroelectric and NQS charge dynamics within a CMOS-compatible materials framework, this work establishes a scalable platform for physical reservoir computing with performance (1000× speed improvement, 5× efficiency gain) that previously required exotic, non-manufacturable material systems. Figure~\ref{fig:FeFET_Reservoir} illustrates a conventional reservoir and its mapping to the proposed FeFET reservoir. Since amorphous oxide semiconductors (AOS) provide multi-modal transport (electron band transport and vacancy hopping), the functionality of interconnected LTM and STM can also be demonstrated with FeFETs having AOS channel, potentially opening the path towards heterogeneously integrated CMOS+X type systems, where X can be overlap engineered FeFETs having AOS channel.

\section{Device Architecture and Characterization}\label{sec2}

\begin{figure}[tbhp]
    \centering
    \includegraphics[width=\linewidth]{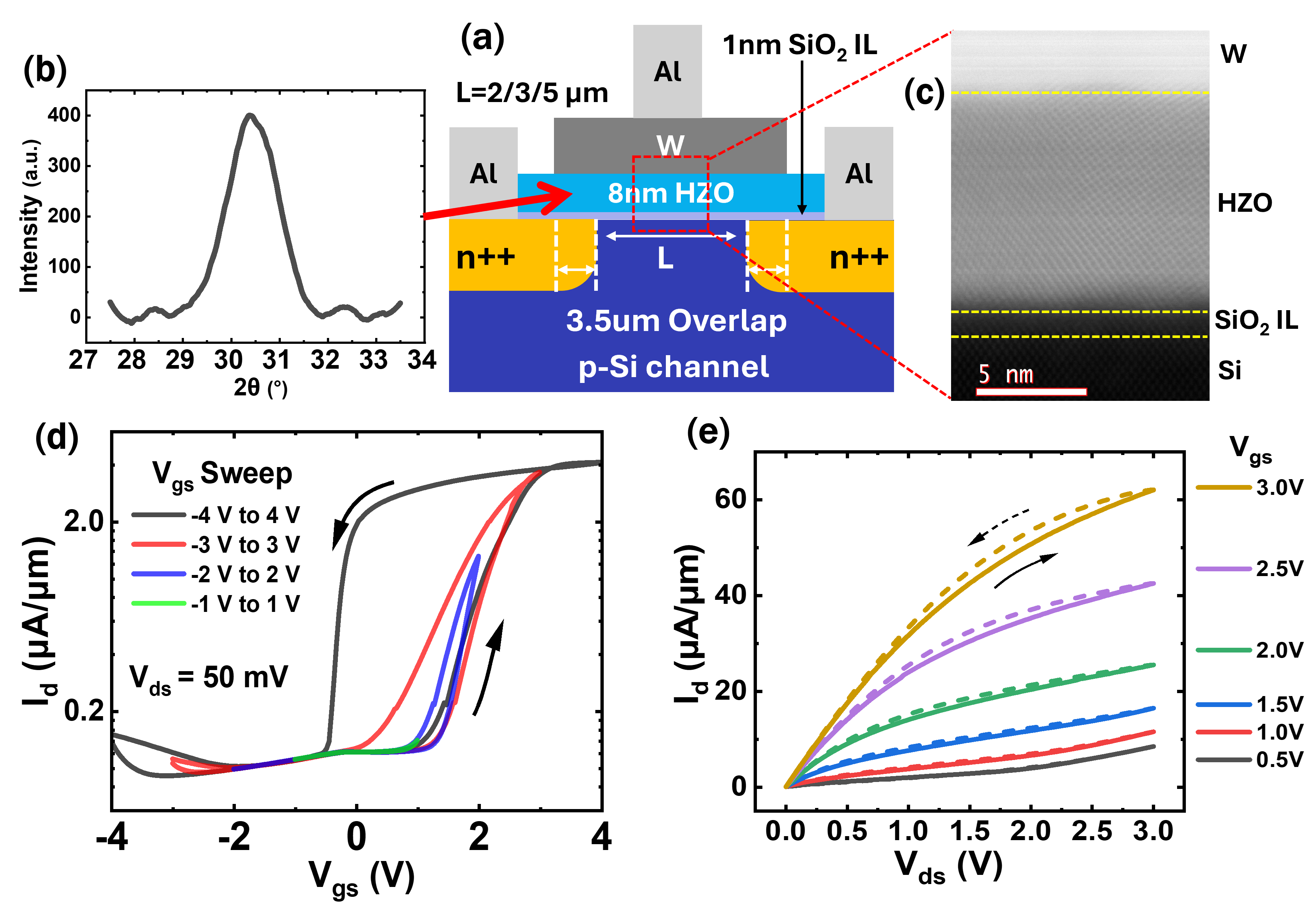}
    \caption{(a) Cross-section of the planar FeFET device showing an overlap region of 3.5 $\mu$m $\times$ 50 $\mu$m between gate metal (W) and source/drain (n++) implant region forming a capacitor structure. (b) X-ray diffraction (XRD) spectrum of the 8 nm hafnium-zirconium-oxide (HZO) film after annealing. The prominent diffraction peak centered at approximately 30.4° corresponds to the (111) reflection of the ferroelectric orthorhombic phase. This confirms the successful crystallization of the HZO layer, which is essential for the ferroelectric behavior of the gate dielectric in the device. (c) A TEM image of the FeFET gate stack. The TEM clearly shows the crystalline nature of the HZO, with the repeating dot-like pattern arising from its polycrystalline structure. (d) $I_{d}$–$V_{gs}$ plot for various gate voltage sweeping ranges showing anticlockwise hysteresis modulating the threshold voltage based on gate polarization. Drain voltage is kept at 50 mV. (e) $I_{d}$–$V_{ds}$ plot for various gate voltages showing anticlockwise hysteresis that is only prominent for faster sweep speed (about 2 V/s) and higher gate voltage. }
    \label{fig:FET}
\end{figure}

The fabricated silicon-based FeFET (Fig.~\ref{fig:FET}a) is based on a conventional CMOS platform, incorporating an 8 nm HZO ferroelectric gate dielectric and a tungsten (W) gate electrode. Fig.~\ref{fig:FET}b gives the XRD analysis plot of the post-annealing HZO. A TEM image of the gate stack is shown in Fig.~\ref{fig:FET}c. The source and drain regions are heavily doped (n++) to ensure low-resistance electrical contacts. Two types of device geometries were implemented in this study: a planar FET and a ring FET (See the image of the fabricated devices in Supplementary Figure 1). The ring FET design offers enhanced electrostatic control. 

Devices were fabricated with channel lengths of 2~\textmu m, 3~\textmu m, and 5~\textmu m. The planar FETs feature a 50~\textmu m channel width, while the ring FETs achieve an effective channel width of approximately 400~\textmu m. This geometric variation enables us to evaluate the FeFET’s memory and computational behavior across multiple structural configurations. A key architectural feature is the deliberate gate–source/drain (G–S/D) overlap, which is critical for the device’s dynamic (short-term) memory response and for the computational capabilities demonstrated here.

Long-term memory (LTM) in FeFETs is controlled by the remanent polarization of the HZO ferroelectric layer. When specific gate voltages are applied, polarization states within HZO are set and remain stable after removing the external gate voltage, yielding robust non-volatile memory. As illustrated in Fig.~\ref{fig:FET}d, the drain-current versus gate-voltage (\(I_{d}\)–\(V_{gs}\)) characteristics exhibit pronounced anticlockwise hysteresis loops (e.g., \(-4\)~V to \(+4\)~V and \(-3\)~V to \(+3\)~V windows), reflecting the existence of distinct and multi-level switchable polarization states, thus justifying the LTM effect.

The overlap capacitance arises from the physical intersection of the gate electrode with the source and drain extensions beyond the channel, forming capacitors at both G–S and G–D interfaces. Its magnitude is governed by the overlap area, dielectric thickness, and the dielectric constant of interfacial layers. \emph{Note that in fast excitation regimes, the overlap is not merely parasitic; it supplies the input coupling that excites a non-quasi-static (NQS) channel charge response.} In an NQS picture, the inversion layer behaves as a distributed, bias-dependent RC line; a rapid drain-voltage perturbation couples through the G–D overlap and perturbs the channel charge near the drain edge, which then relaxes with a finite time constant \(\tau\).

Accordingly, when voltage pulses (or sufficiently fast \(V_{ds}\) sweeps) are applied, the device exhibits \emph{transient}, history-dependent changes in \(I_{d}\) that appear as apparent hysteresis in the drain-current versus drain-voltage (\(I_{d}\)–\(V_{ds}\)) characteristics (see Fig.~\ref{fig:FET}e). We attribute this behavior to NQS channel-charge relaxation rather than to quasi-static conduction changes: the short-term memory (STM) is volatile and decays as the NQS charge perturbation relaxes, yielding paired-pulse facilitation/depression (PPF/PPD) on microsecond timescales. Consistent with compact NQS theory \cite{OhWardDutton1980}, the dominant time constant scales approximately as
\[
\tau \;\sim\; R_{\mathrm{ch}}(V_{gs},V_{ds},L,W)\;C_{\mathrm{ov}}(\text{overlap geometry},t_{\mathrm{ox}}),
\]
so that shorter \(L\) and operation deeper in strong inversion (lower \(R_{\mathrm{ch}}\)) reduce \(\tau\), whereas operation near/under threshold (higher effective \(R_{\mathrm{ch}}\)) increases \(\tau\). The overlap geometry and dielectric stack set \(C_{\mathrm{ov}}\), providing a separate design knob for tuning \(\tau\). This engineered overlap—a designed material functionality rather than a parasitic effect—enables the microsecond-scale dynamics essential for real-time edge computing.

The interplay between overlap-driven NQS STM and ferroelectric LTM enables the coexistence of volatile and non-volatile memory within a single FeFET device. This dual-memory modality enhances the computational richness of the reservoir in three ways: (i) the overlap-driven NQS dynamics introduce nonlinear, history-dependent current responses, expanding the diversity of internal states; (ii) STM encodes and processes recent input sequences essential for temporal tasks; and (iii) LTM (via polarization) configures the operating regime (strong vs.\ weak inversion) and thereby programmatically tunes the sign and strength of STM (PPF vs.\ PPD) under a fixed \(V_{gs}\). Together, fast, volatile STM and slow, persistent LTM project input signals into a high-dimensional feature space suitable for reservoir computing and low-power edge inference.

\begin{figure}[tbhp]
    \centering
    \includegraphics[width=1\linewidth]{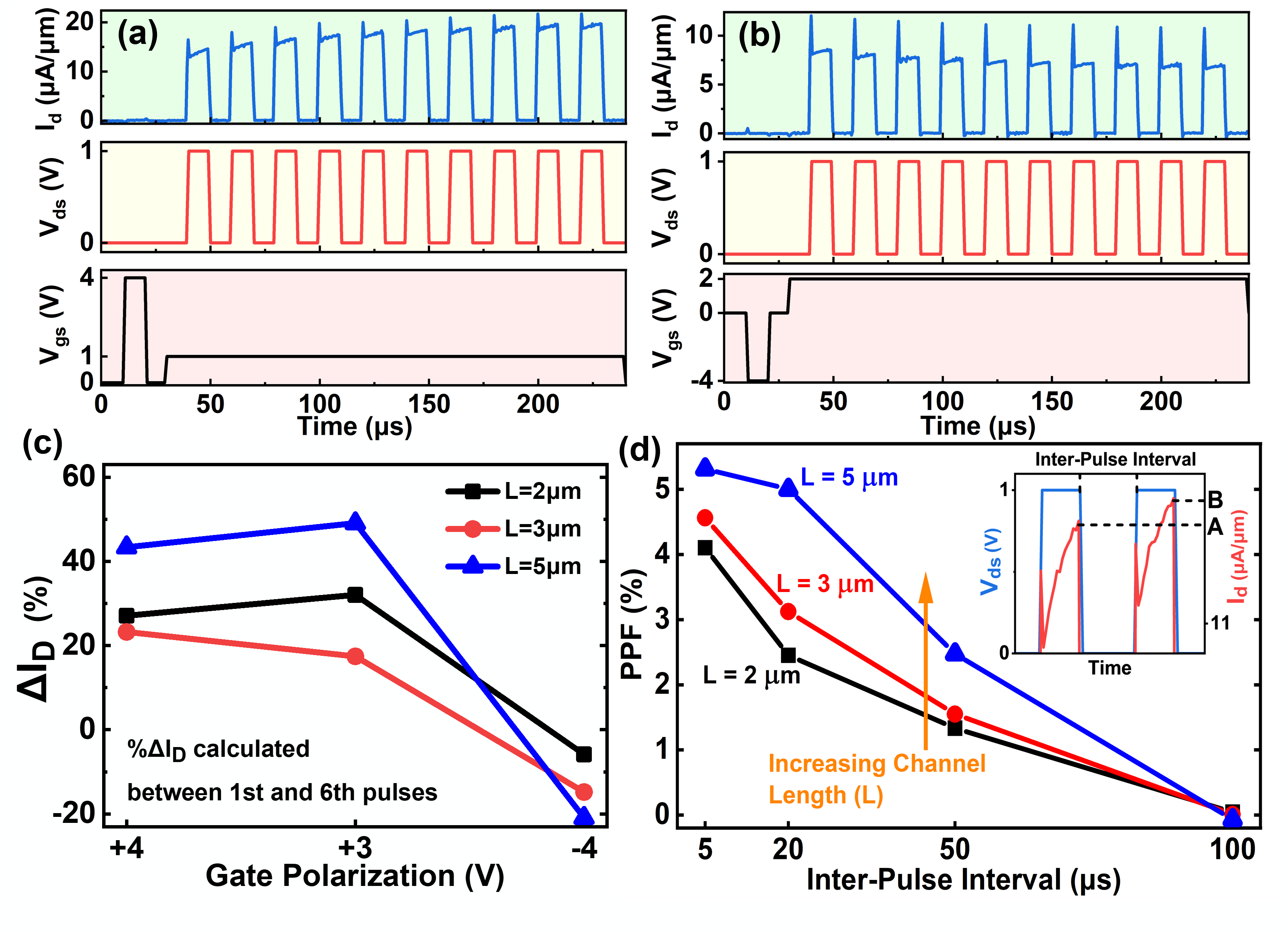}
    \caption{(a) Current response to ten 10 $\mu$s rectangular voltage pulses of 1 V administered in close temporal proximity, separated by an inter-pulse interval of 10 $\mu$s. The pulses generate an amplified current response, illustrating the PPF of the FeFETs. Drain voltage pulse sequence showing PPF effect when the FeFET is polarized at +4 V on the gate side. (b) Drain voltage pulse sequence showing PPD effect when the FeFET is polarized at -4 V on the gate side. (c) Calculated percentage increase of drain current between the first and sixth pulses during the ten-pulse sequence. The percentage shows a positive increase in drain current for +4 V and +3 V gate polarization, while a decrease in drain current occurs at -4 V gate polarization. (d) To demonstrate the unique temporal dynamics of each FeFET device, we display representative experimental data for PPF for three planar devices with various channel lengths of 2, 3, and 5 $\mu$m. The inset displays a typical drain current response to two square voltage pulses administered in close temporal proximity, separated by an inter-pulse interval (IPI). The pulses generate an amplified current response, illustrating the PPF of the FeFET. Quantitatively, PPF is defined in terms of the peak drain current of the first pulse (A) and second pulse (B) as follows: PPF = (B - A)/A × 100\%. The PPF values for the three planar devices of our FeFETs are then plotted against IPI. The voltage pulses employed were 15 $\mu$s in duration and rectangular waveforms with an amplitude of 1 V.}
    \label{fig:PPF}
\end{figure}

\section{Ferroelectric-Modulated Transient Charge Dynamics: Inter-dependence between Short-term and Long-term Memory}\label{sec3}

To demonstrate the interaction between STM and LTM in our silicon-based FeFETs, we performed systematic electrical measurements using a series of drain voltage pulse sequences following controlled gate polarization. This approach reveals how the device’s remanent polarization state—set by the gate voltage—modulates the transient current response, enabling tunable memory dynamics essential for reservoir computing.

After polarizing the FeFET gate to a specific voltage (either +4 V, +3 V, or –4 V), we applied a sequence of ten 10~\textmu s rectangular drain voltage pulses (1 V amplitude), each separated by a 10~\textmu s inter-pulse interval. The resulting drain current (I$_{d}$) was recorded in real time to capture the device’s dynamic response to each pulse. This protocol was repeated on planar FETs with varying channel lengths, and the results are consistent, enabling a comprehensive analysis of memory effects with respect to the polarization states.

The drain current response shows that the device’s gate polarization state directly controls the strength and nature of its short-term memory. When the device is polarized at +4 V or +3 V, the drain current response to the pulse sequence exhibits a paired pulse facilitation (PPF) effect. Here, successive pulses generate incrementally larger current peaks, indicating that the device retains and amplifies the memory of recent inputs. This is observed in Figure~\ref{fig:PPF}a, where the current increases with each pulse during the sequence. In contrast, polarizing the gate at –4 V induces a paired pulse depression (PPD) effect. The drain current response diminishes with each subsequent pulse, reflecting a rapid decay of the STM trace. Figure~\ref{fig:PPF}b highlights this behavior, where the current response decreases over the pulse train. A summarized Fig.~\ref{fig:PPF}c shows the calculated percentage increase in drain current against different polarization states. To further quantify and analyze this interplay between gate polarization and short-term memory, we developed a dedicated “read–write–read” testing protocol. The results, presented in Supplementary Figures 2–4, clearly demonstrate that the percentage change in drain current between read pulses systematically varies with the gate polarization state, confirming that long-term polarization modulates short-term memory strength and dynamics across all device geometries.

The physical origin of these tunable STM effects—PPF and PPD—is understood to arise from two interconnected mechanisms: a dominant non–quasi–static (NQS) channel-charge relaxation excited via the gate–source/drain overlap capacitance, and a secondary, minor contribution from charge trapping/detrapping. The device's long-term memory state, set by the ferroelectric gate polarization, directly modulates this NQS transient response.

The primary mechanism is the overlap-induced capacitive coupling that perturbs channel charge near the drain edge and relaxes with a finite NQS time constant. When the gate is positively polarized (+4 V or +3 V at our fixed $V_G$), the local threshold is lowered, and the device operates in \emph{strong inversion}; a drain pulse leaves a small residual pre-charge at the channel edge if the inter-pulse interval is shorter than the NQS time constant, leading to higher initial conductance on the next pulse (PPF). Conversely, with negative polarization (–4 V), the operating point shifts toward \emph{weak inversion}/near-threshold; the relaxation is slower and diffusion-dominated, so a residual deficit persists at the second pulse, reducing the current (PPD). The proof that operating in weak or strong inversion dictates PPD or PPF is shown in Supplementary Figure 5. In this case, the positive polarization demonstrates PPF when operated in strong inversion during drain pulsing, and negative polarization demonstrates PPD when operated in weak inversion. This signifies that the operating regime (strong vs. weak inversion) determines the type of short-term memory, which can be modulated using gate polarization during training, providing reconfigurability to the RC network for different training datasets.

A secondary, less dominant mechanism involves the trapping and detrapping of electrons into and out of defect states, either within the HZO gate oxide or at its interfaces. This process could explain a small portion of the hysteresis observed in slow-sweep DC I$_d$–V$_{ds}$ characteristics (Fig.~\ref{fig:FET}e). However, our simulation results suggest that the trapping/detrapping is quite small for our pulsed experiments~\cite{kim_fast_2017,ma_defects_2022}. Additionally, given that the device operates in a relatively low-voltage regime that limits carrier injection into the oxide, the contribution from trapping is considered minimal. Therefore, the fast pulsed response is dominated by the overlap-engineered NQS mechanism programmed by the non-volatile ferroelectric state.

The existence and dynamic nature of short-term memory in the silicon-based FeFET devices are demonstrated in Figure~\ref{fig:PPF}d. This plot presents representative paired-pulse facilitation (PPF) data from three planar FeFET devices with channel lengths of 2, 3, and 5~\textmu m. In each case, two consecutive square voltage pulses are delivered to the device with varying inter-pulse intervals (IPI). The figure reveals a typical drain current response: upon applying two closely timed voltage pulses, the second pulse generates a noticeably higher current than the first, indicating the presence of short-term facilitation. The percentage PPF is calculated as: 
\begin{equation}
    \label{eq:PPF}
    \mathrm{PPF} = \frac{(B - A)}{A} \times 100\%.
\end{equation}

A key observation from Fig.~\ref{fig:PPF}d is that the PPF effect diminishes gradually as the interval between pulses increases. This decay in PPF with increasing IPI is consistently observed across all three device geometries. Such an exponential-like decay with IPI is characteristic of NQS relaxation governed by the channel time constant $\tau$ (see Supplementary Note). The clear reduction in facilitation at longer inter-pulse intervals directly evidences the volatile, time-dependent nature of short-term memory encoded in the device response. These temporal dynamics are crucial for enabling temporal processing in reservoir computing, as they provide fading memory essential for sequence-dependent computation.

These results confirm that our FeFETs’ long-term memory, encoded by the stable remanent polarization of the HZO layer, can control the STM strength and sign. By switching the gate polarization, we can reversibly toggle between facilitation and depression, effectively tuning the device’s temporal response characteristics. These findings prove that our FeFETs possess robust LTM (via ferroelectric polarization) and highly tunable STM (via overlap-driven NQS dynamics coupled with the polarization controlled operating regime). The ability to modulate STM strength and sign through long-term polarization states significantly increases the heterogeneity of accessible reservoir states. This dual-memory mechanism enhances the computational richness and dynamic range of the reservoir, enabling more effective temporal information processing—a key requirement for advanced neuromorphic and reservoir computing applications. The CMOS-compatible materials foundation—exploiting established ferroelectric HZO and silicon processing—provides an immediately manufacturable pathway to scalable neuromorphic systems. In contrast, many reported memristive reservoir systems rely on exotic materials that are incompatible with CMOS processing, hindering practical deployment.

\begin{figure}[tbh]
    \includegraphics[width=1\linewidth]{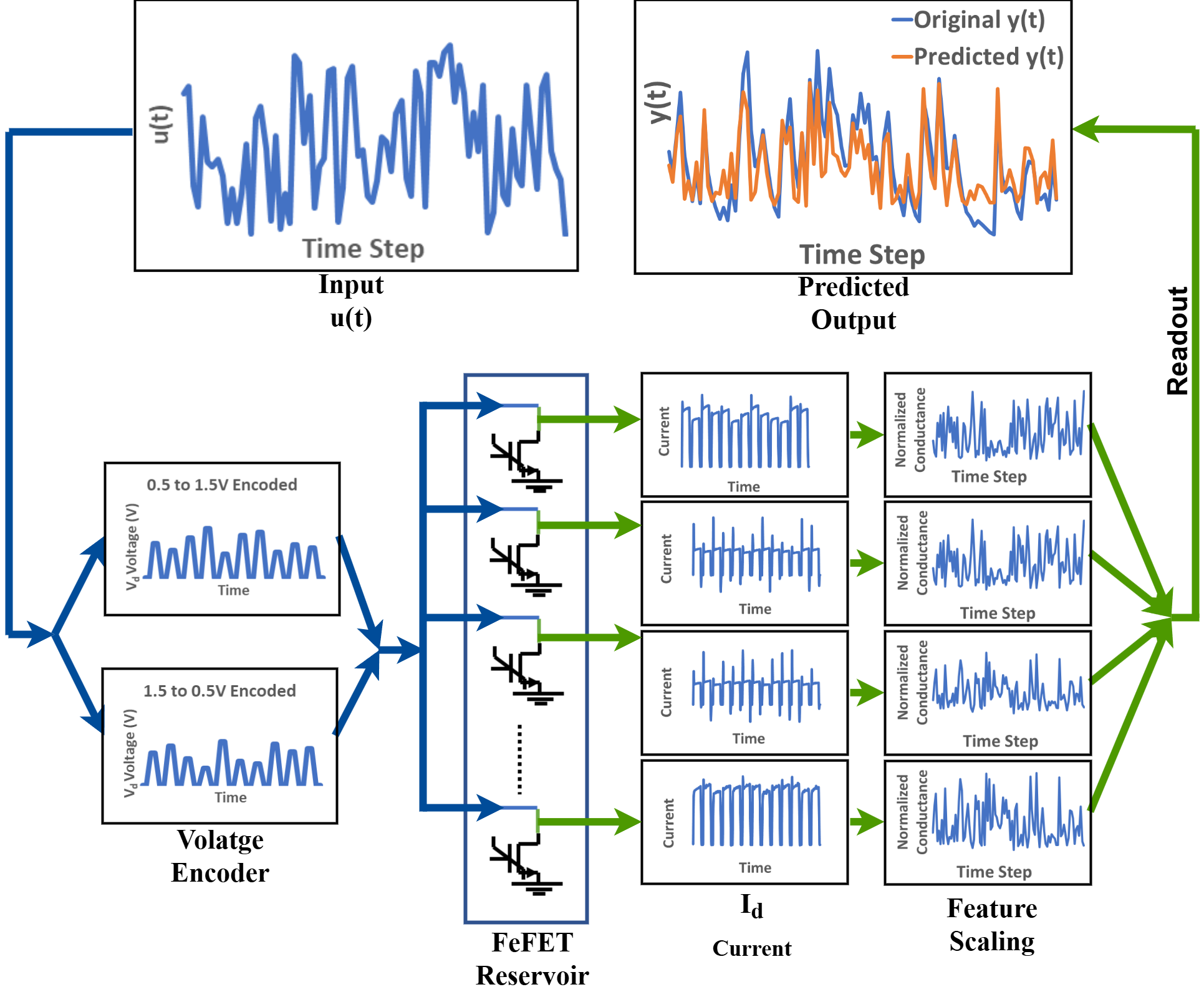}
    \caption{Complete workflow solving the second-order dynamic nonlinear transfer function with our Si-FeFET devices. }
    \label{fig:ProblemSolvingProcessFlow}
\end{figure}

\section{FeFET-based Physical Reservoir Computing Implementation}\label{sec4}

To implement physical reservoir computing with our silicon-based FeFET devices, we used the FeFET reservoir to predict a second-order dynamic nonlinear transfer function. The transfer function is described as follows:
\begin{equation}
    \label{eq:2nd_non_linear}
    y(t) = 0.4y(t-1) + 0.4y(t-1) y(t-2) +0.6u^3(t) +0.1
\end{equation}

The output signal, $y(t)$, depends on the present input, $u(t)$, as well as the previous two inputs, $y(t-1)$ and $y(t-2)$ (that is, a time delay of two time steps), as shown in Equation \ref{eq:2nd_non_linear}. In this study, we trained the FeFET-based RC system to map a random input onto a higher-dimensional space. This mapping enables the generation of an accurate second-order dynamic nonlinear transfer function output from the input after training, without prior knowledge of the underlying mathematical relationship between input and output. The process flow of how the whole system works is demonstrated in Figure~\ref{fig:ProblemSolvingProcessFlow}.

\begin{figure}[tbhp]
    \centering
    \begin{subfigure}[b]{0.49\textwidth}
        \includegraphics[width=\linewidth]{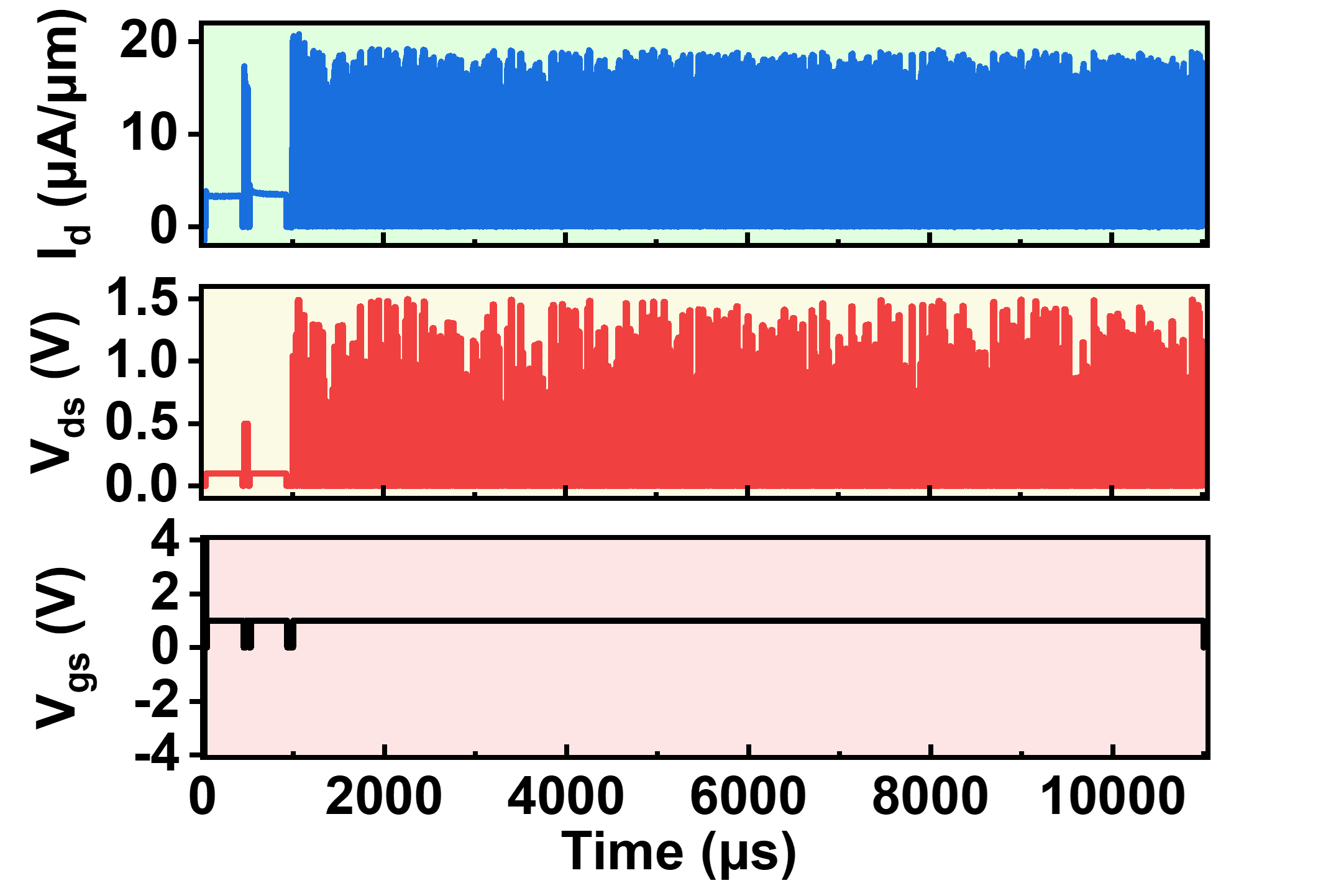}
        \caption{}
        \label{fig:500pa}
    \end{subfigure}
    \hfill
    \begin{subfigure}[b]{0.49\textwidth}
        \includegraphics[width=\linewidth]{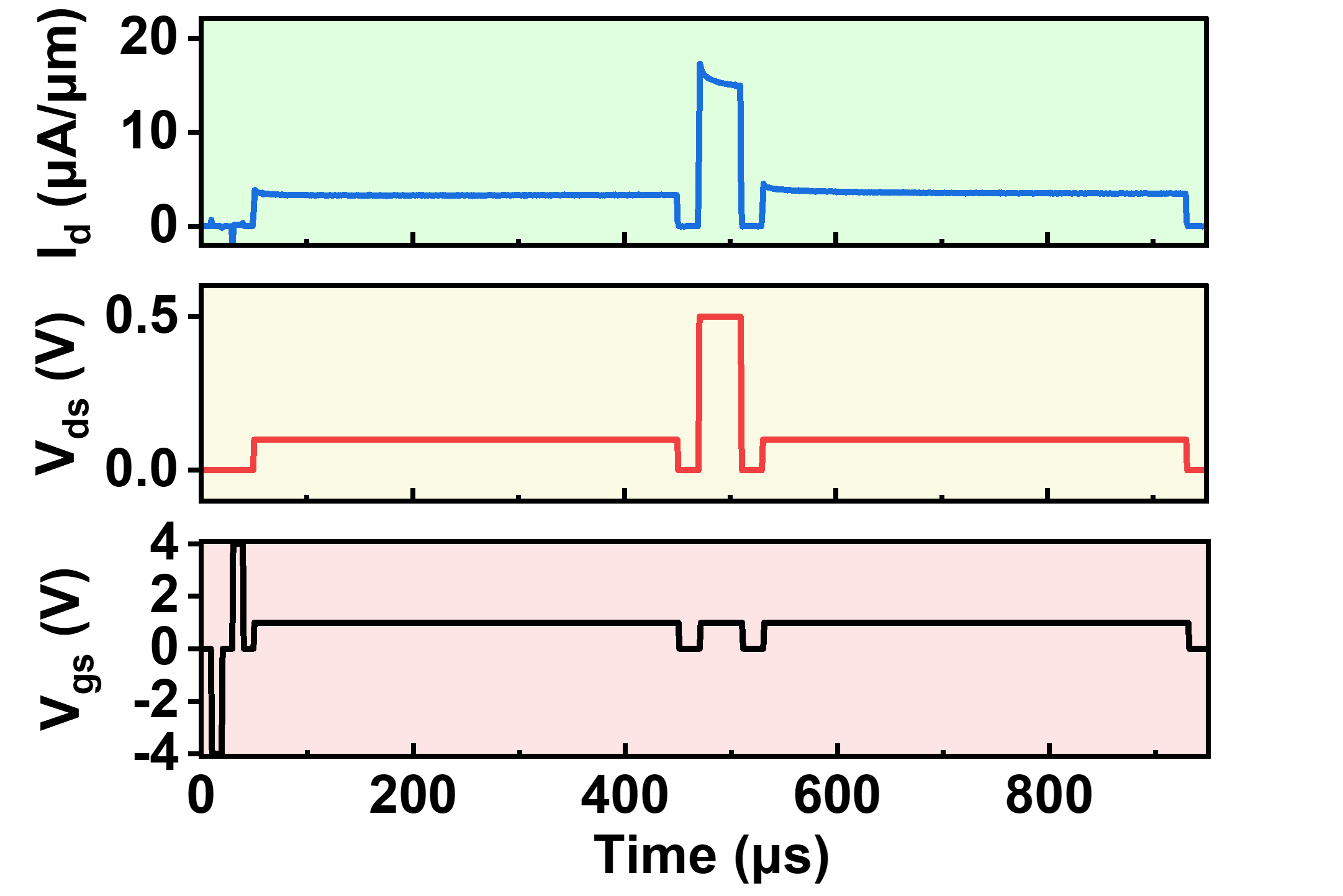}
        \caption{}
        \label{fig:500pb}
    \end{subfigure}
    \begin{subfigure}[b]{0.53\textwidth}
        \includegraphics[width=\linewidth]{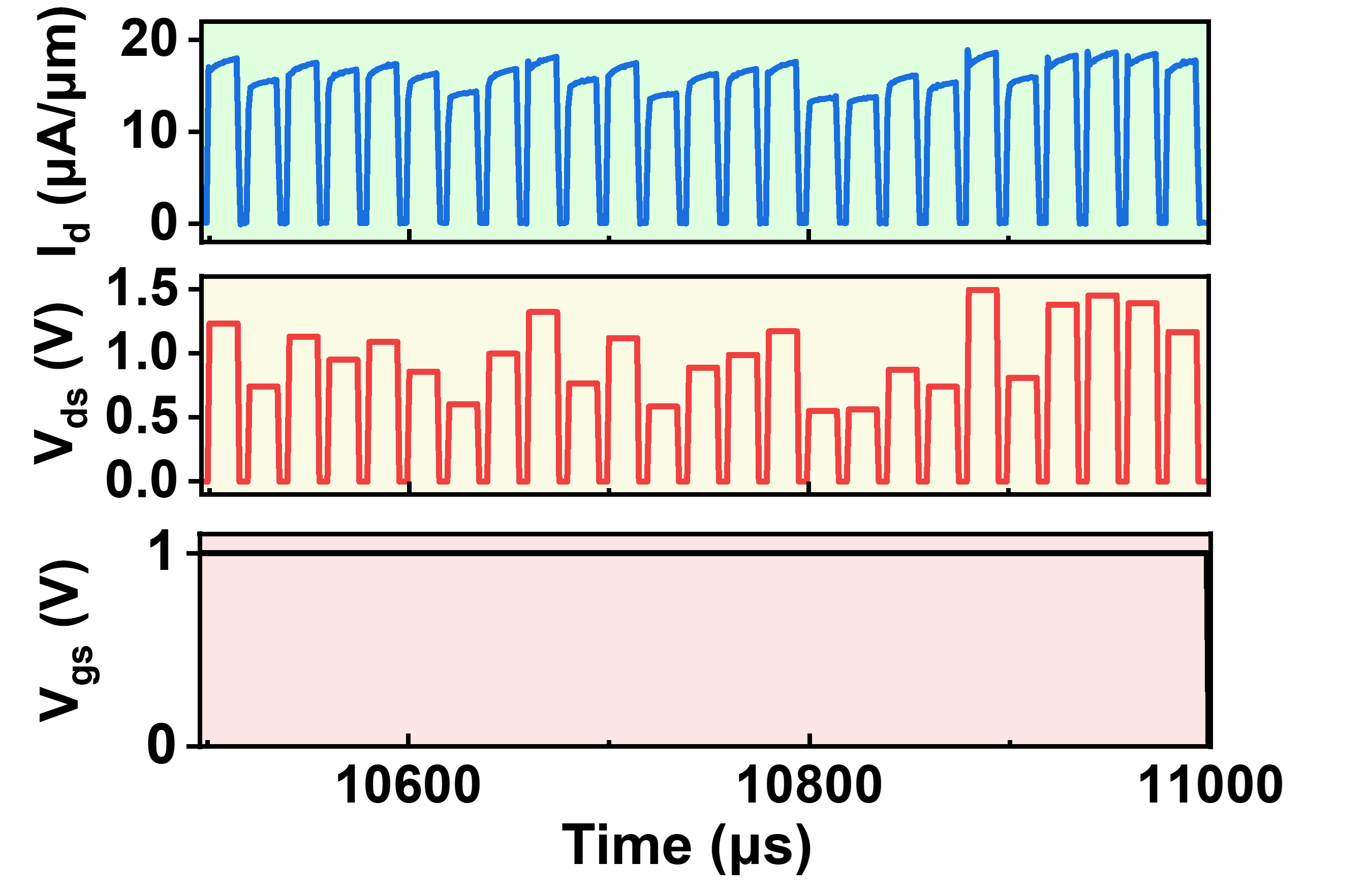}
        \caption{}
        \label{fig:500pc}
    \end{subfigure}
    \begin{subfigure}[b]{0.45\textwidth}
        \includegraphics[width=\linewidth]{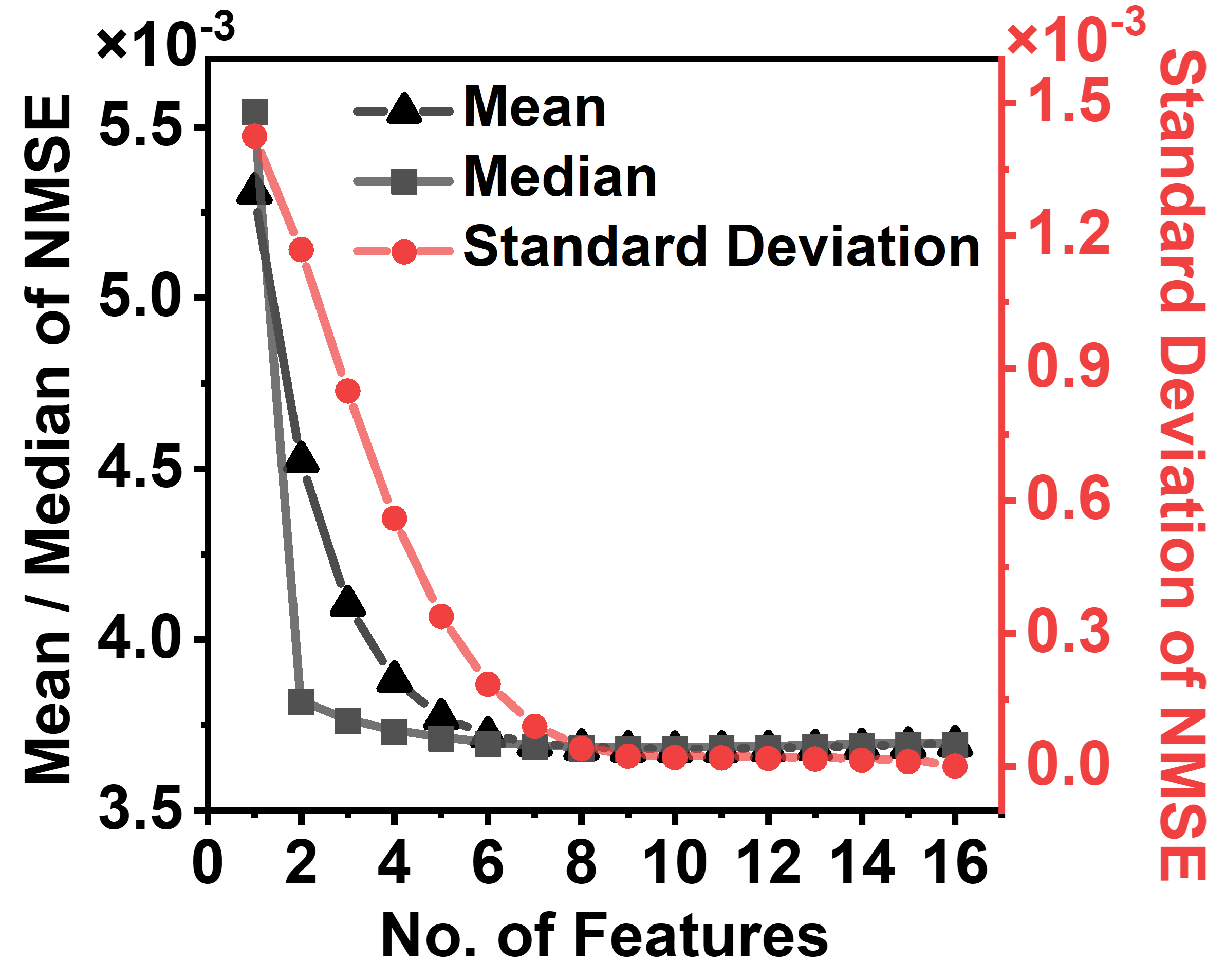}
        \caption{}
        \label{fig:statistical_NMSE}
    \end{subfigure}
    \caption{(a) Complete timeline of the 500-point training set measurements. (b) The pre-writing scheme of the 500-point training set measurements. The pre-writing scheme is utilized to account for the initialization of the device.  (c) Section of the encoded voltage pulses and drain current response towards the end of the 500-point training set measurements. (d) Due to device heterogeneity, each FeFET feature in the reservoir exhibits different polarization states, and different feature combinations result in varying NMSE values. The plot illustrates how the statistical measures of NMSE change with the number of features, showing a clear decrease in NMSE as the number of features increases. }
    \label{fig:500p}
\end{figure}

We encode both training and testing sequences as a series of voltage input pulses delivered to the drain terminal. Specifically, the input data, ranging from 0 to 0.5, is mapped onto a sequence of rectangular drain voltage pulses linearly ranging from 0.5 V to 1.5 V. Each pulse has a period of 20 $\mu$s and a 75\% duty cycle. As visualized in the attached experimental setup (Fig.~\ref{fig:500p}a), each measurement cycle begins with a designed pre-writing sequence to account for device initialization effects. This pre-writing (Fig.~\ref{fig:500p}b) consists of gate ferroelectric polarization set pulses followed by a drain ‘read, write, read’ voltage sequence. This protocol stabilizes the ferroelectric polarization state and ensures consistent device behavior across repeated measurements. Supplementary Figure 4 provides more details on the pre-writing scheme. 

Once initialized, the device is fed by a stream of 500 encoded drain voltage pulses, as shown partially in Fig.~\ref{fig:500p}c. Each pulse in this sequence represents a time point in the input signal, allowing the reservoir to process dynamic, time-dependent patterns. This pulse train is configured to maximize the device’s nonlinearity and memory properties, which are critical for effective reservoir computing. During every voltage pulse, the corresponding drain current response is recorded and utilized as the instantaneous output of the physical reservoir. This time series of drain current values reflects the evolving internal state of the FeFET reservoir in response to the input sequence, capturing both short-term memory (capacitive) and long-term memory (ferroelectric polarization) contributions. These experimentally acquired current responses for each pulse form the feature set that defines the reservoir’s output layer.

After pulse stimulation, this data is passed to the reservoir computing readout layer, the only trained portion of the system, where machine learning algorithms extract and map relevant patterns for the target computational task. The readout layer consists of the same number of neurons as the reservoir, with one extra for bias. Before going through the readout layer, the reservoir output current is first converted into conductance and then normalized with the first conductance sample of the set. The normalized conductance is then scaled between 0 and 1, depending on the minimum and maximum normalized conductance for each feature. For the training and testing sets, both conductance normalization and feature scaling are performed with reference to the training set. In the readout layer, Ridge Regression~\cite{hastie_ridgereg_2017} is used for training. 

\begin{figure}[tbhp]
    \centering
    \includegraphics[width=0.8\linewidth]{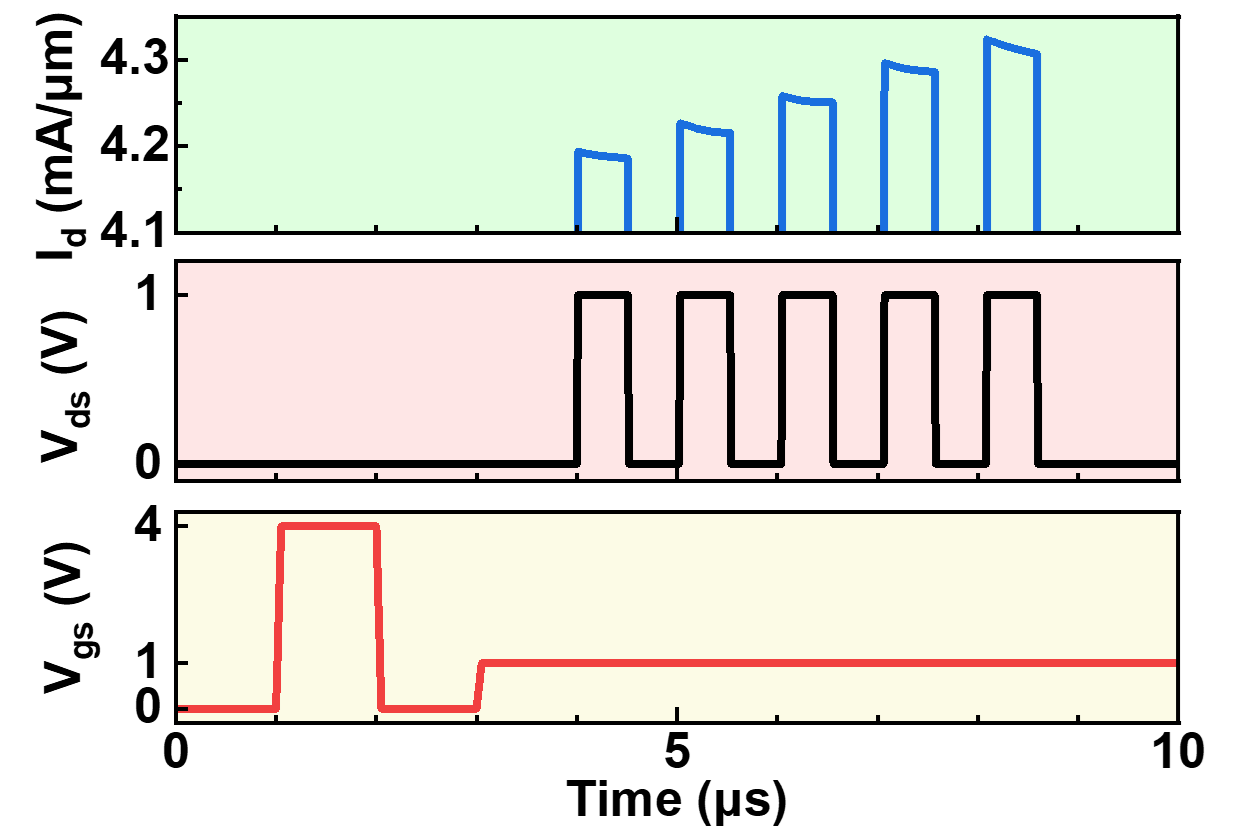}
    \caption{Simulated drain current (I$_d$) as a function of time is shown with respect to the corresponding drain voltage (V$_{ds}$) and gate voltage (V$_{gs}$) pulsing schemes. An initial gate voltage pulse from 0 V to +4 V and back to 0 V precedes the final set at +1 V. This sets the positive polarization at the gate. Five drain pulses are encoded between 0 V and +1 V with a duty cycle of 50\% and a pulse period of 1 $\mu$s. The resulting I$_d$ shows clear PPF.}
    \label{fig:simPPF}
\end{figure}

\section{Results and Discussion}\label{sec12}
When evaluating performance in comparison with other published results, several key metrics highlight the advantages of our proposed FeFET-based reservoir computing systems. In terms of prediction accuracy, our FeFET approach demonstrates a competitive normalized mean square error (NMSE) of $3.31 \times 10^{-3}$ for training and $3.69 \times 10^{-3}$ for testing tasks. Most importantly, our system achieves this performance having only 16 reservoir states, $>3\times$ reduction from Hossain \emph{et al.} and $>5\times$ reduction from Du \emph{et al.}, indicating a highly compact and efficient design. Also, this reduction in the number of reservoir states results in a similar reduction in the size of the readout layer, providing significant scalability to our system. Furthermore, our FeFET-based reservoir consumes only $1.50 \times 10^{-7}$~J of energy while offering three orders of magnitude ($\approx10^3\times$) faster response time of 20~$\mu$s compared to Hossain \emph{et al.} and Du \emph{et al.}, which is essential for CMOS integration and large-scale deployment. Additionally, the FeFET-based system benefits from device-level heterogeneity through tunable parameters such as gate polarization, gate voltage, and device geometry; an important feature missing in prior works that enriches its computational dynamics without compromising scalability. Fig. \ref{fig:statistical_NMSE} shows the statistical change in NMSE with different combinations of FeFET heterogeneity. Our results suggest that at least 8 features are sufficient to reach close to the lowest error. Further increases in the number of features improve the stability of the result by reducing the standard deviation of the errors for different possible feature combinations. Supplementary Fig. 6 provides more details on FeFET heterogeneity in the context of error stability. Collectively, these attributes establish FeFET-based reservoir computing as a scalable, fast, energy-efficient, and CMOS-compatible platform capable of supporting complex temporal processing with a compact footprint. Tables \ref{tab:comparison_metrics_1} and \ref{tab:comparison_metrics_2} summarize the comparison with recently published work in the literature. Beyond performance metrics, the CMOS-compatible material platform enables heterogeneously integrated CMOS+X type device arrays — a scalability pathway absent in material systems requiring high-temperature deposition or unconventional substrates.

{
\renewcommand{\arraystretch}{1.5} % Adjust row height
\begin{table}[h]
    \centering
    \begin{tabularx}{\textwidth}{|c|X|X|X|X|X|}
        \hline
        Work & Train [NMSE] & Test [NMSE] & Reservoir states & Reservoir energy [J] & Response Time \\ \hline
        This work & $3.31 \times 10^{-3}$ & $3.69 \times 10^{-3}$ & 16 & $1.5 \times 10^{-7}$ & 20 µs  \\ \hline
        Hossain et al.\cite{najem_bio_memcap_2023} Work & $5.75 \times 10^{-4}$ & $7.81 \times 10^{-4}$ & 50 & $2.72 \times 10^{-8}$ & 600 ms  \\ \hline
        Du et al.\cite{du_reservoir_2017} Work & $3.61 \times 10^{-3}$ & $3.43 \times 10^{-3}$ & 90 & $3.34 \times 10^{-4}$ & 20 ms  \\ \hline
    \end{tabularx}
    \caption{Comparison of key metrics between some of the notable reported results.}
    \label{tab:comparison_metrics_1}
\end{table}
}

{
\renewcommand{\arraystretch}{1.5} % Adjust the value as needed
\begin{table}[h]
    \centering
    \begin{tabularx}{\textwidth}{|c|X|X|X|X|X|}
        \hline
        Work & CMOS Compatibility & Memory Type & Readout Layer Size & Device Level Heterogeneity & Scalability \\ \hline
        This work & Yes & Both Long \& Short-Term & 17 & Yes\footnotemark & Yes \\ \hline
        Hossain et al.\cite{najem_bio_memcap_2023} Work & No & Short-Term & 51 & No & No \\ \hline
        Du et al.\cite{du_reservoir_2017} Work & Yes & Short-Term & 91 & No & Yes \\ \hline
    \end{tabularx}
    \footnotetext[1]{Gate Polarization, Gate Voltage, Device Geometry.}
    \caption{Comparison of key metrics between some of the notable reported results (cont...).}
    \label{tab:comparison_metrics_2}
\end{table}
}

Takagi \emph{et al.} \cite{takagi_physical_2023} reported similar work that addresses the second-order nonlinear dynamic task using a FeFET-based reservoir computing architecture. However, the formulation of the problem reported in that paper diverges significantly from our approach. Specifically, instead of sampling the input values in Equation~\ref{eq:2nd_non_linear} from a continuous uniform distribution over the interval $[0, 0.5]$, the input is restricted to discrete values, namely $\{0, 0.5\}$. Due to this fundamental difference in input encoding and task formulation, the results published in Takagi \emph{et al.} are not directly comparable and were therefore excluded from the quantitative evaluation presented in this work.

Material-level scalability, determined by whether ferroelectric-semiconductor coupling maintains functionality at advanced nodes, is critical for practical deployment. To investigate the scalability of our device, we modeled a silicon (Si) FeFET using the Ginestra\texttrademark{} EDA simulation software with the same device structure but scaled to 30 nm gate length. The simulated electrical characteristics show good agreement with the experimental data, with minor deviations arising from geometrical differences between the fabricated and modeled devices; these differences were introduced to improve simulation convergence and speed. The simulated device structure is shown in Supplementary Figure~7 and its parameters are detailed in the Methods section.

The simulated drain current (I$_d$) in response to the pulsed drain voltage (V$_{ds}$) is presented in Figure~\ref{fig:simPPF}. Before the drain pulse, positive polarization at the gate is set by applying a gate voltage (V$_{gs}$) of +4~V. During the simulations, the source and bulk electrodes were grounded, V$_{gs}$ was set to +1~V to turn the transistor on, and the temperature was held at 300~K. The results clearly exhibit paired-pulse facilitation (PPF). In our interpretation, these transients reflect a \emph{non-quasi-static (NQS)} channel-charge response: drain-voltage steps couple through the gate--source/drain overlap capacitance, perturbing the inversion charge, which then relaxes with a finite time constant rather than instantaneously as assumed in a quasi-static picture. In order to demonstrate the decaying nature of this memory, simulations were run with a longer inter-pulse interval (IPI) between subsequent V$_{ds}$ pulses. As seen in Supplementary Figure~8, after the first pulse, the current amplitude does not change with successive pulses, confirming the presence of STM in the modeled device. This fading behavior with IPI is consistent with an NQS relaxation process of the channel charge. Additionally, there is an initialization effect (increase in I$_d$) only after the first pulse to set the baseline; an effect that was also observed in the experimental device (Supplementary Figure~4). This strong agreement between simulation and experimental results validates the employed physics models, affirming their suitability for further scalability studies.

Simulation result analysis identifies the overlap capacitance between the gate and the source/drain regions as the primary driver of this NQS STM. Mechanistically, the overlap capacitance provides the input coupling while the inversion channel behaves as a distributed, bias-dependent RC; together they generate a non-quasi-static (NQS) “charge-deficit” relaxation with a characteristic time constant governed by the channel input resistance and oxide capacitance, \cite{BSIM4_manual,Cheng_NQS,OhWardDutton1980,PaulosAntoniadis1983} yielding microsecond-scale short-term memory. Under our fixed gate-bias condition, the ferroelectric polarization selects the local operating regime at the drain edge: positive polarization lowers the effective threshold and places the device in strong inversion (shorter NQS time constant), whereas negative polarization shifts the operating point toward weak inversion (longer NQS time constant). Consistent with this NQS bias dependence, \cite{BSIM4_manual,Cheng_NQS} we observe paired-pulse facilitation in the strong-inversion case and paired-pulse depression in the weak-inversion case. This overlap-engineered NQS mechanism is highly scalable, as reducing the transistor area allows us to operate at a higher frequency regime, which may not always be possible for memristor based short-term memory. As detailed in Supplementary Figure~9, after an initial +4~V gate pulse sets the device’s polarization, the ferroelectric state remains largely stable during subsequent drain pulses, making its contribution to PPF/PPD minimal. Similarly, the simulation shows negligible changes in the trapped charge-carrier density, effectively ruling it out as a significant contributor to STM. With these other effects discounted, the analysis confirms that the overlap-capacitance-driven NQS channel response is the dominant mechanism behind the observed STM. Therefore, engineering the gate overlap remains the source of the STM as the device is scaled down, proving the scalability of our device structure.

It is important to note that scaling down the transistor also scales the overlap area, which reduces the overlap capacitance. In order to keep the capacitance the same, the thickness of the dielectric must also scale. Equivalently, from an NQS standpoint the characteristic time constant scales approximately with the product of channel resistance and overlap capacitance (schematically, $\tau \!\sim\! R_{\mathrm{ch}} C_{\mathrm{ov}}$), \cite{BSIM4_manual} so $\tau$ can be maintained or shortened by co-optimizing geometry/materials (to set $C_{\mathrm{ov}}$) and bias/length (which set $R_{\mathrm{ch}}$). Additionally, the pulse width can be scaled, thereby proportionally scaling the system-level response time, provided that the edge (or pulse) duration remains on the order of, or shorter than, $\tau$, so that the NQS fading-memory behavior is preserved.

\section{Conclusion}\label{sec14}

In this work, we have successfully demonstrated a CMOS-compatible ferroelectric-semiconductor platform featuring an engineered gate–source/drain (G–S/D) overlap capacitance that can be programmed to exhibit non–quasi–static (NQS) short-term memory (STM) tunable via gate polarization switching. With this engineered material architecture, we have demonstrated a highly efficient and tunable physical reservoir computing platform. By systematically characterizing and leveraging the device's unique dual-memory modalities, we have demonstrated its capacity to solve a second-order nonlinear dynamic task with high accuracy, achieving a competitive test NMSE of $3.69 \times 10^{-3}$ using only 16 reservoir states. This performance highlights the computational capability of this highly compact and efficient design.

The core of our approach lies in the synergistic interplay between ferroelectric long-term memory (LTM) and overlap-capacitance–engineered NQS short-term memory (STM). We established that LTM, governed by the remanent polarization of the HZO gate dielectric, acts as a non-volatile control for the device's transient dynamics. By setting the gate polarization, we could deterministically switch the device's response between paired-pulse facilitation (PPF) and paired-pulse depression (PPD), effectively tuning the STM strength. This tunable STM, which originates from the engineered overlap capacitance between the gate and source/drain regions, provides the rich, nonlinear, and fading memory essential for reservoir computing.

Beyond demonstrating reservoir computing performance, this work establishes material-engineering principles for neuromorphic hardware: (i) dual-timescale functionality through orthogonal mechanisms (ferroelectric switching + capacitive dynamics), (ii) material-parameter tunability (overlap capacitance, polarization state, operating regime) enabling application-specific optimization, and (iii) CMOS-compatible material choices (HZO, silicon, standard processing) ensuring manufacturability. These principles suggest broader applicability: the ferroelectric-modulated charge dynamics demonstrated here may make it possible to design nanosecond to millisecond tunable volatile memory using oxide semiconductors. This can ultimately enable adaptive analog circuits, reconfigurable sensors, and materials-based signal processing beyond neuromorphic computing.

The CMOS-compatible fabrication, low energy consumption ($1.5 \times 10^{-7}$~J), fast response time (20~\textmu s), and potential scalability position this ferroelectric-semiconductor platform as an excellent candidate for next-generation edge computing applications. Although our current devices have a fixed overlap geometry, our simulation results indicate that this structural parameter is a critical design knob for optimizing NQS-driven STM. This suggests a clear path for future work in co-designing device physics and reservoir performance by further exploring the gate–S/D overlap and enabling multi-modal transport using oxide semiconductors. Ultimately, this research establishes that incorporating both LTM and overlap-driven NQS STM within a single CMOS-compatible material system  provides a powerful, scalable, and energy-efficient pathway for realizing advanced neuromorphic systems for complex temporal data processing.

\section{Methods}\label{sec15}
\subsection{Device fabrication}\label{subsec:fab}

The fabrication of the silicon-based FeFET begins with the etching of alignment marks on a silicon wafer. A screen oxide is thermally grown at 950 \textsuperscript{o}C for 30 minutes to prepare for source and drain implantation. Ion implantation follows, involving photolithographic patterning, n++ implantation, and photoresist stripping. The screen oxide is then removed using a buffered oxide etch (BOE). Subsequently, the implanted dopants in the source and drain regions are activated through annealing at 900 \textsuperscript{o}C for 20 minutes.

A standard CMOS RCA clean is performed before gate oxide formation. During the SC1 step, approximately 1 nm interfacial SiO$_{2}$ layer is chemically grown on the silicon surface. The gate stack, consisting of an 8 nm ferroelectric Hf$_{0.5}$Zr$_{0.5}$O$_2$ (HZO) layer, is deposited using thermal atomic layer deposition (ALD) at 250 \textsuperscript{o}C. A 40 nm tungsten (W) gate electrode is then deposited, followed by rapid thermal annealing at 500 \textsuperscript{o}C for 30 seconds to crystallize the ferroelectric film. The tungsten is patterned and etched to form the gate electrode, which is extended over the doped source and drain regions to create the desired overlap of 3.5 $\mu$m. 

Next, the gate oxide is selectively etched to remove the HZO from the source and drain areas. A 400 nm SiO$_{2}$ interlayer dielectric is deposited and patterned to define contact vias using dry etching. Finally, aluminum pads (containing 1\% silicon) are deposited via a lift-off process, and the device undergoes a forming gas anneal at 400 \textsuperscript{o}C for 30 minutes to improve contact integrity and overall reliability.

\subsection{Electrical measurements}\label{subsec:meas}

Device electrical measurements were performed at room temperature under standard laboratory lighting. A Micromanipulator\textsuperscript{\textregistered} P200L probe station was used to establish contact with the devices. The station was connected to a Keysight\textsuperscript{\textregistered} B1500A Semiconductor Device Analyzer mainframe, which served as the primary instrument for characterization. For dynamic characterization, a Keysight\textsuperscript{\textregistered} B1530A Waveform Generator/Fast Measurement Unit (WGFMU) module was utilized. This specialized module generated the necessary voltage pulses applied to the gate and drain terminals for the pulsed current-voltage (I-V) measurements while concurrently measuring the resulting drain current.

\subsection{Computational Modeling}\label{subsec:sim}

Device simulations were performed using the Ginestra$^{TM}$ software package, a trap-centric modeling platform that self-consistently describes all the relevant physical mechanisms occurring in semiconductors, dielectric layers, and novel materials. The simulated device was based on a conventional n-type MOSFET structure with a physical gate length of 30 nm and an effective channel length of approximately 6 nm (calculated by excluding the overlap regions).

\textbf{Device Structure and Parameters:} The simulated structure consisted of a 200 nm thick silicon substrate with a width of 200 nm and a total length of 100 nm. The substrate was uniformly doped with a p-type doping concentration of $2 \times 10^{15}$ $cm^{-3}$. The source and drain regions were defined with a length of 30 nm each and doped with an n-type concentration of $1 \times 10^{19}$ $cm^{-3}$, which extends under the gate stack. A 5 nm spacer separated the gate from the source and drain contacts. In order to maintain consistency with the experiments, the gate stack was composed of an 8 nm thick ferroelectric Hf$_{0.5}$Zr$_{0.5}$O$_2$ (HZO) layer on top of a 1 nm SiO$_{2}$ interfacial layer, capped with a tungsten (W) metal gate. An ideal bulk electrode at the bottom of the substrate and ideal source/drain electrodes on the top surface were used, with the bulk contact held at 0 V bias.

\textbf{Physics and Material Models:} The ferroelectric behavior of the HZO layer was modeled using the Landau-Khalatnikov formalism combined with the Ginzburg domain coupling term, available within Ginestra$^{TM}$~\cite{Sharma2022IRPS}. The accurate modeling of the interaction between atomic defects and FE properties was ensured by the coupled solution with Poisson's equation, while considering a uniform oxygen vacancy distribution with a concentration of $5 \times 10^{17}$ $cm^{-3}$ in the HZO layer. Electron and hole transport models considering intrinsic (direct/Fowler-Nordheim tunneling, thermionic emission, drift) and defect-assisted mechanisms were implemented in the framework of the multi-phonon trap-assisted tunneling (TAT) theory.

\backmatter

%\bmhead{Supplementary information}

\pdfbookmark[1]{Acknowledgements}{acknowledgements}\bmhead{Acknowledgements}
\addcontentsline{toc}{section}{Acknowledgements}
Simulation was performed using Ginestra$^{TM}$, an Applied Materials proprietary simulation software platform. More information can be accessed at \cite{ginestra}. The authors acknowledge the suggestion and guidance of Valerio Lunardelli, Luca Larcher, and Gaurav Thareja from Applied Materials, Inc. throughout the project. R.I. also acknowledges the help and support of Purdue University College of Engineering. 

\section*{Declarations}

\begin{itemize}
\item \textbf{Funding}: Y.W. and M.S.S. acknowledge the support by the U.S. National Science Foundation under Standard Grant No. 2521468 and 2521469, respectively. N.V. acknowledges the support by the U.S. Nuclear Regulatory Commission Faculty Development Grant. A. P. acknowledges the FAR 2024 project of the “Enzo Ferrari” Engineering department of the University of Modena and Reggio Emilia, Italy, for the financial support.

\item \textbf{Consent for publication} All authors certify that the work has been reviewed by each author and all authors are consenting for the publication if accepted.

\item \textbf{Data and Code availability} All data and code are fully available upon reasonable request to the corresponding authors.
\item \textbf{Author contributions}: Y.W., M.S.H and R.I. conceived the idea. Y. W. did all the electrical characterization, and M.S.S. performed the reservoir computing training and tests based on the data provided by Y.W. S.S. and L.F. participated in the device fabrication, N.V., A.P. contributed to the simulation of the fabricated device using Ginestra$^{TM}$. A.I.K., M.S.H., and R.I. provided the overall supervision of the work. All the authors participated in manuscript writing and proofreading.   

\item \textbf{Competing interests}: All the authors declare no competing interests.

\end{itemize}

%\noindent
%If any of the sections are not relevant to your manuscript, please include the heading and write `Not applicable' for that section. 

%\bigskip

%\bigskip\noindent
%Springer journals and proceedings: \url{https://www.springer.com/gp/editorial-policies}

%\bigskip\noindent
%Nature Portfolio journals: %\url{https://www.nature.com/nature-research/editorial-policies}

%\bigskip\noindent
%\textit{Scientific Reports}: \url{https://www.nature.com/srep/journal-policies/editorial-policies}

%\bigskip\noindent
%BMC journals: \url{https://www.biomedcentral.com/getpublished/editorial-policies}
%\end{flushleft}

\begin{appendices}

\section{Supplementary information}\label{secA1}

\begin{figure}[H]
    \includegraphics[width=\linewidth]{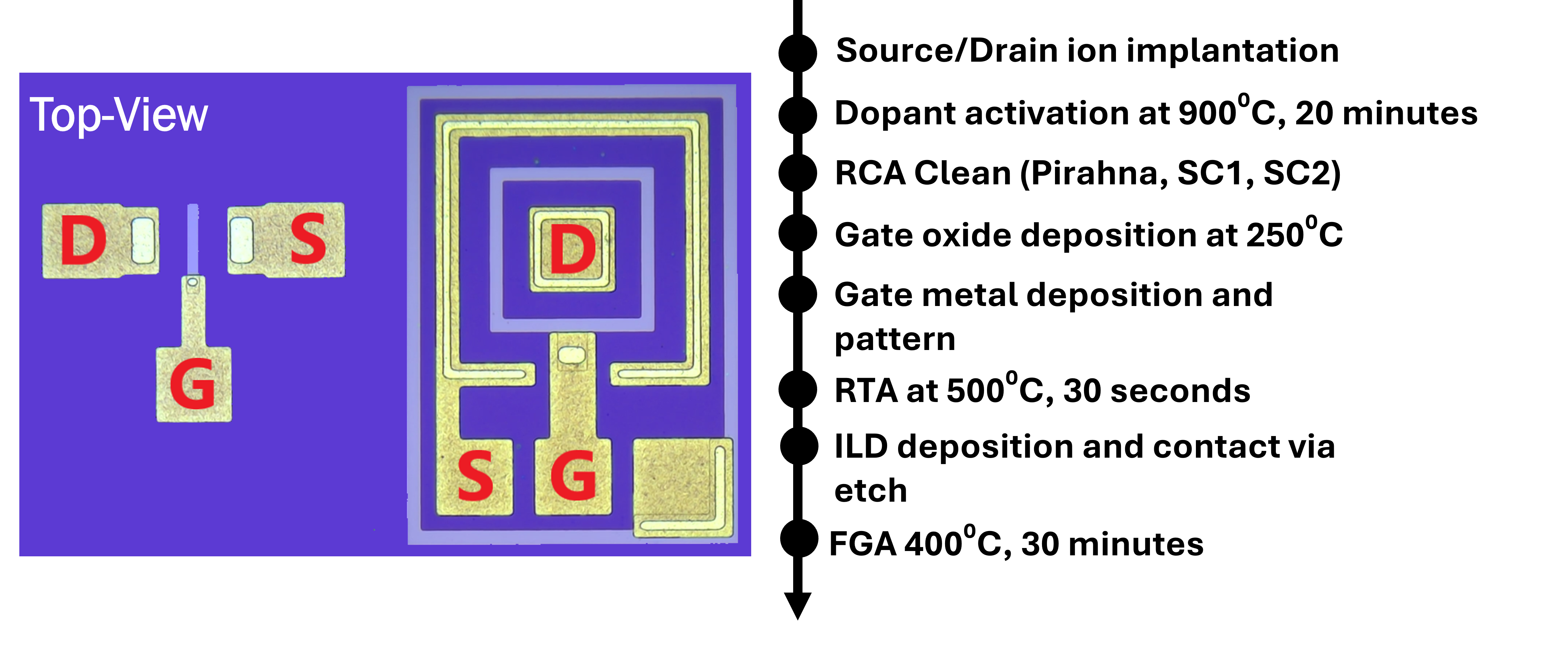}
    \caption*{\textbf{Supplementary Figure 1 \texttt{|} Device top-view and fabrication process.}
    The image on the left side shows the top view of fabricated devices. Devices were made in two different geometries: planar (left) and ring (right) FETs. The flowchart on the right side illustrates the device's fabrication process flow. }
    \label{fig:deviceFab}
\end{figure}

\begin{figure}[H]
    \includegraphics[width=\linewidth]{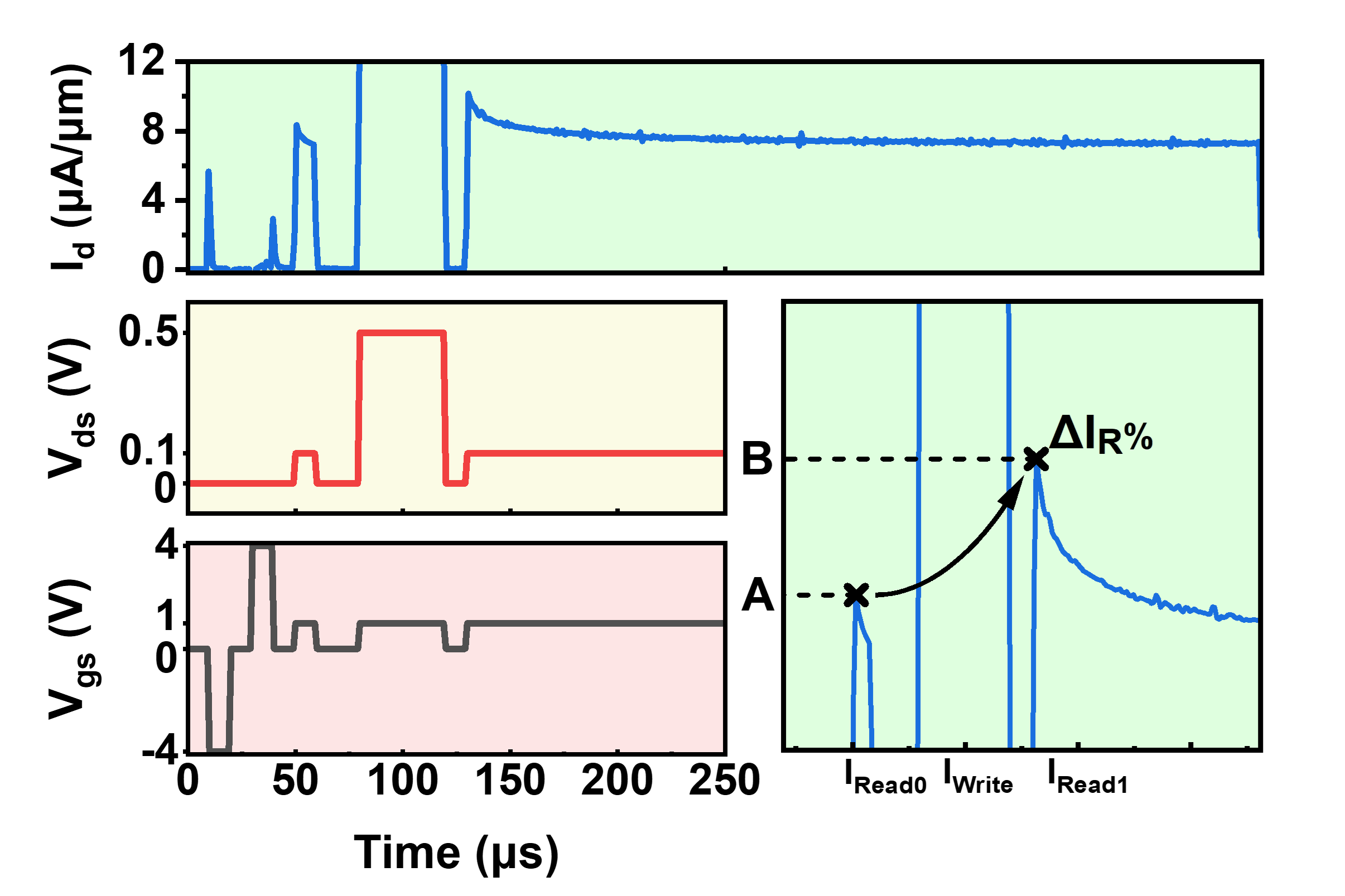}
    \caption*{\textbf{Supplementary Figure 2 \texttt{|} Short-term memory strength measurement.}
    The device is given a -4V, 10 $\mu$s gate voltage pulse followed by a specific gate voltage for 10 $\mu$s to clear and set the desired gate polarization. Then we apply a series of drain voltage pulses of 0.1 V, 0.5 V, and 0.1 V as the read0, write1, and read1 pulses to test the effect of the drain write pulse. The calculated percentage change between the two read currents (as shown in the lower right sub-figure) is then plotted against different gate polarizations in the next supplementary figure. }
    \label{fig:STMread}
\end{figure}

\begin{figure}[H]
    \centering
    \includegraphics[width=0.8\linewidth]{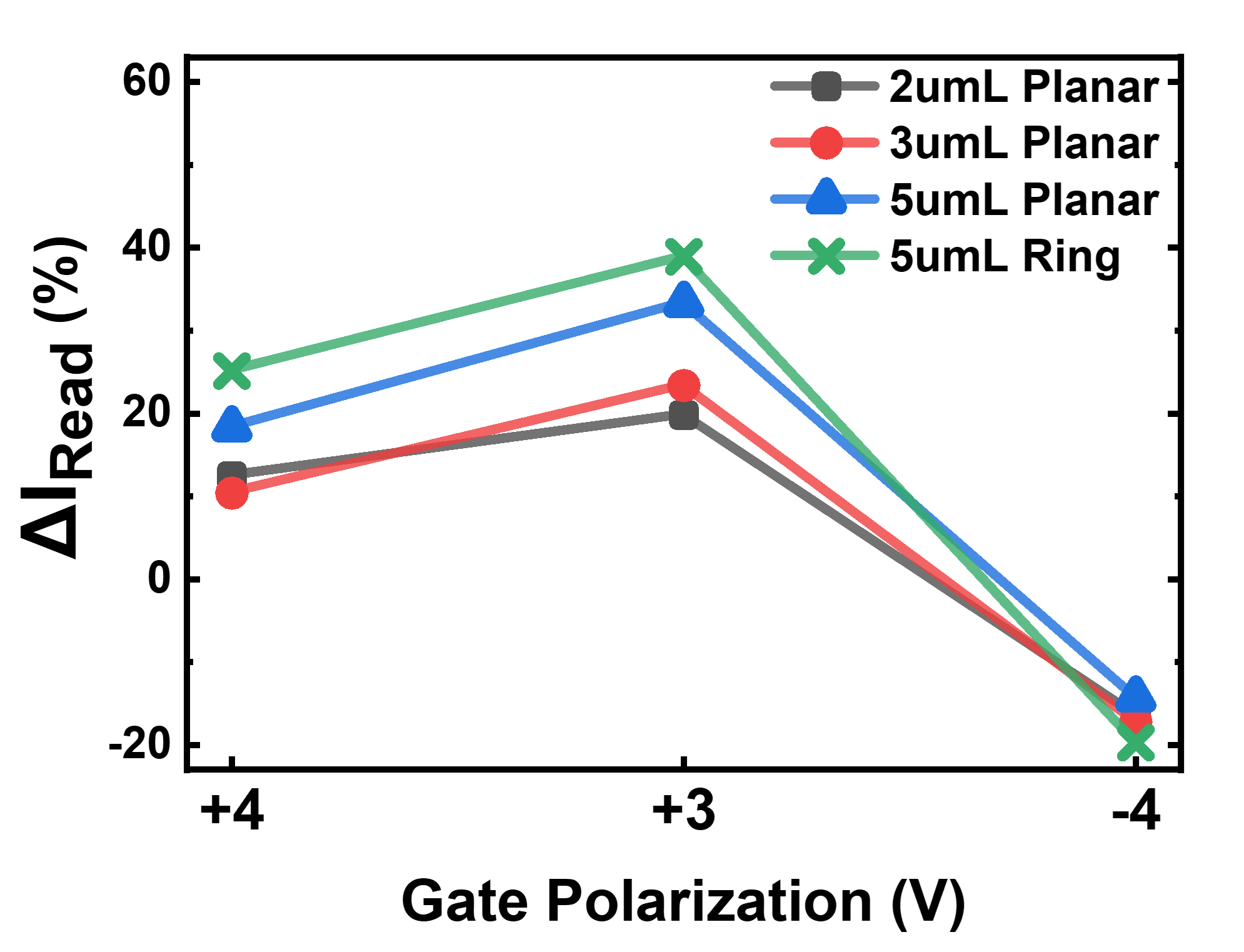}
    \caption*{\textbf{Supplementary Figure 3 \texttt{|} Short-term memory strength affected by long-term polarization.}
    This plot shows the percentage change in drain read current against different gate polarizations. Four lines represent all four geometries of our FeFETs used. The drain current response shows that the device’s gate polarization state directly controls the strength and nature of its short-term memory. When the device is polarized at +4 V or +3 V, the second drain read current increases compared to the baseline $I_{d}$ read current. Polarizing the gate at –4 V and then holding the gate voltage at 2V during the drain voltage pulse causes the second $I_{d}$ read current to decrease. This figure demonstrates how the device's long-term memory state modulates its short-term memory response.}
    \label{fig:STMpercent}
\end{figure}

\begin{figure}[H]
    \includegraphics[width=\linewidth]{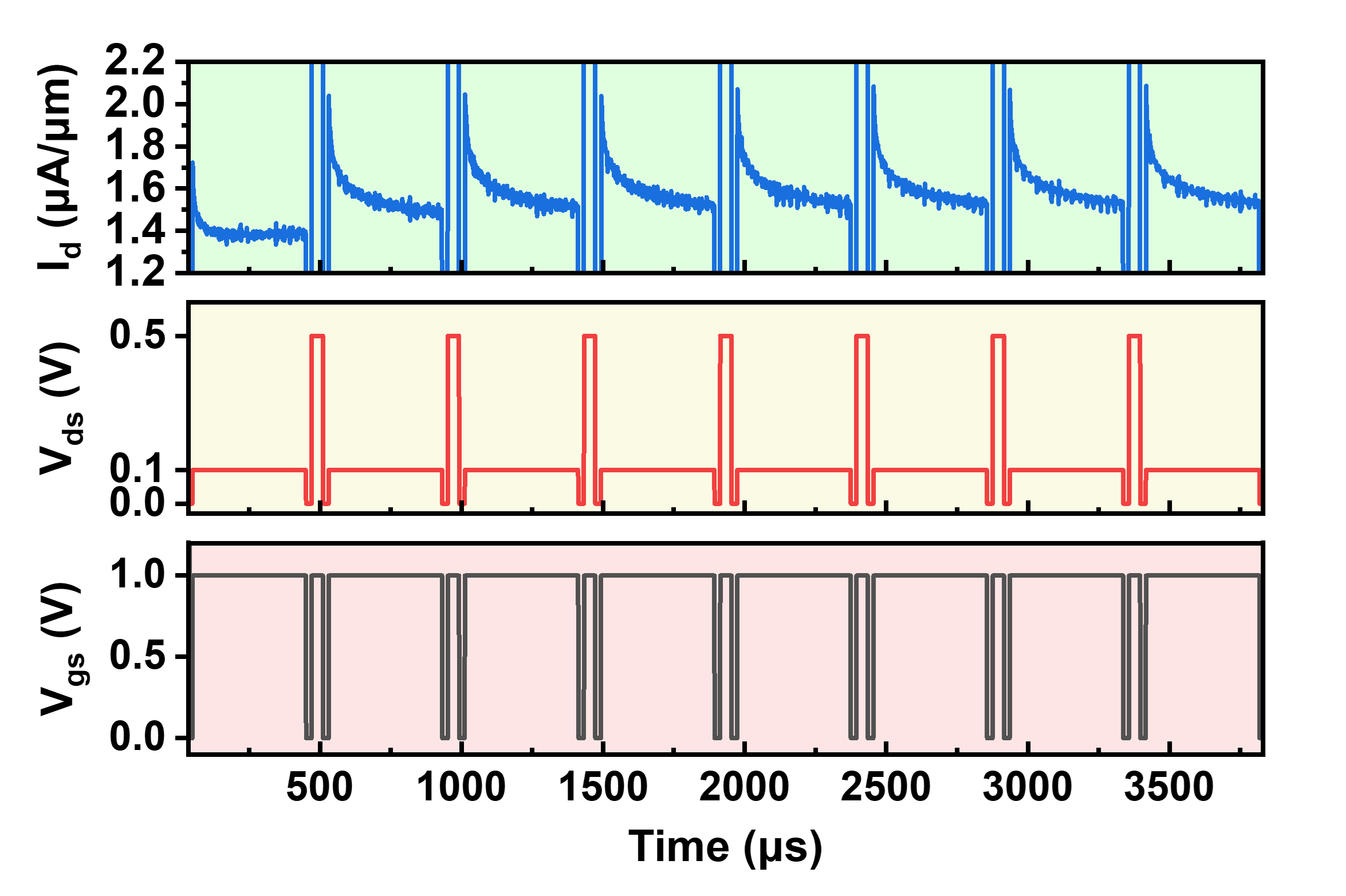}
    \caption*{\textbf{Supplementary Figure 4 \texttt{|} Multiple short-term write and read sequence.}
    The device is gate polarized at +4 V. Then, the drain terminal inputs a baseline read voltage, followed by seven write and read voltage pulse sequences. The drain current response shows some initialization effect from the baseline read to the second read pulse. After that, the $I_{d}$ write and read current responses are consistent. Based on this, we use the first three voltage pulses (read, write, read) as the pre-writing scheme to set up the device at the start of the reservoir computing electrical measurement. }
    \label{fig:STM}
\end{figure}

\begin{figure}[H]
    \centering\includegraphics[width=0.8\linewidth]{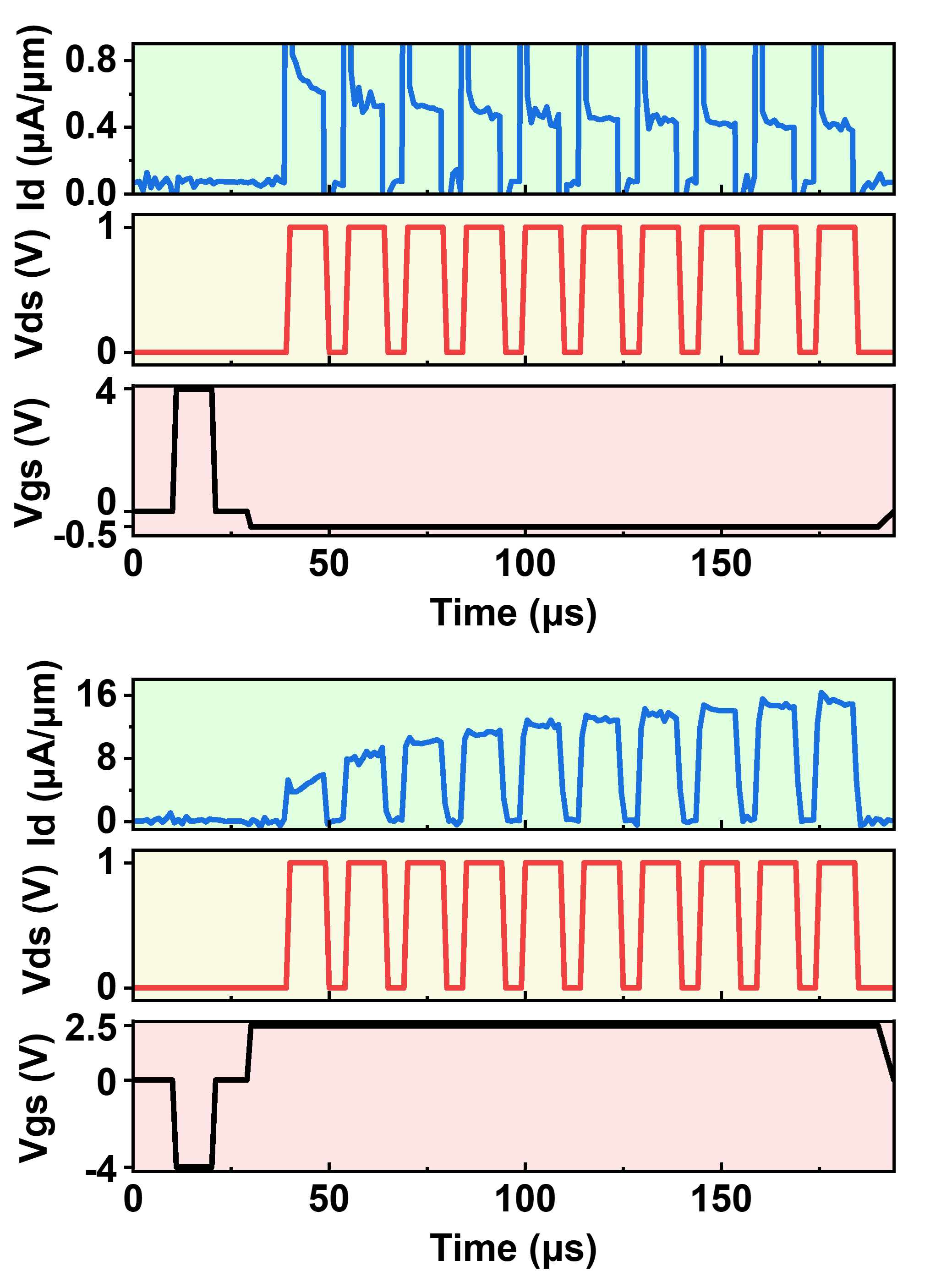}
    \caption*{\textbf{Supplementary Figure 5 \texttt{|} Verification of the non-quasi-static (NQS) origin of the observed short-term memory (STM).}
    In the upper plot, for a positively polarized device, applying a small negative gate bias (-0.5 V) during the drain‑pulse train places the channel in weak inversion and yields paired‑pulse depression (PPD) in drain current under identical pulses. This is aligned with Figure~\ref{fig:PPF}b, where applying a weak inversion of +2 V gate bias for a negatively polarized device creates PPD. In the lower plot, for a negatively polarized state, holding V$_{gs}$ at +2.5 V during the same V$_{ds}$ pulse sequence drives strong inversion and produces paired‑pulse facilitation (PPF). This is aligned with Figure~\ref{fig:PPF}a, where applying a strong inversion of +1 V gate bias for a positively polarized device creates PPF. These two plots show expected drain current short-term memory consistent with a non‑quasi‑static (NQS) channel‑charge response. 
    }
    \label{fig:STM}
\end{figure}

\begin{figure}[H]
    \centering\includegraphics[width=1.0\linewidth]{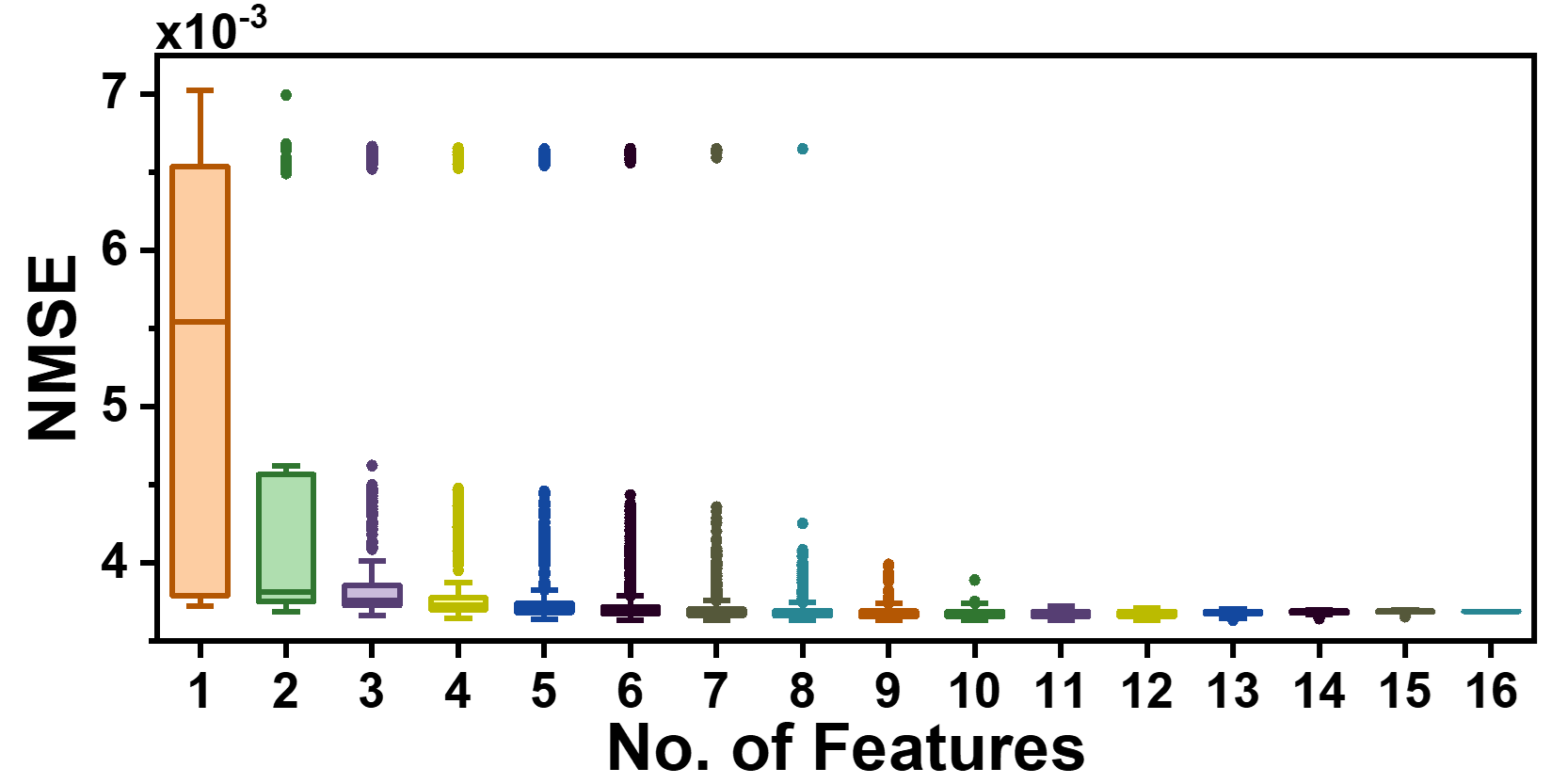}
    \caption*{\textbf{Supplementary Figure 6 \texttt{|} Change of NMSE due to Heterogeneous Reservoir.}
    In our experiment, we have 16 different heterogeneous reservoirs with different gate polarization, gate voltage, voltage encoding, and device geometry. The box plot shows the change of NMSE with different combinations of reservoirs. We can see that after 11 features, all metrics shrank into a small range and are showing a consistent NMSE result.}
    \label{fig:BoxPlotNMSE}
\end{figure}

\begin{figure}[H]
    \includegraphics[width=\linewidth]{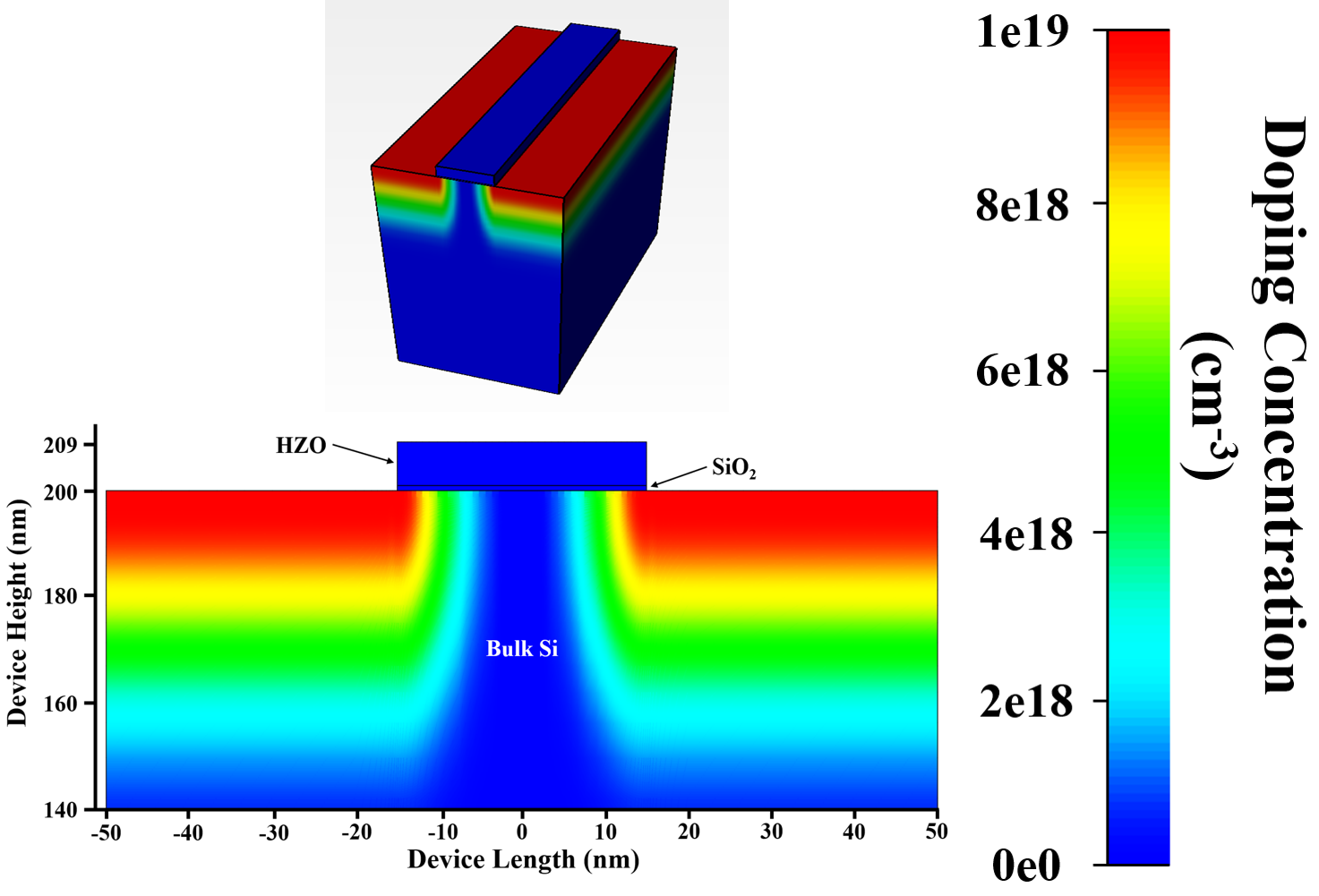}
    \caption*{\textbf{Supplementary Figure 7 \texttt{|} Geometry and doping profile of simulated device visualization.}
    A visualization of the device setup, colormapped by doping concentration. The cross-section of the 3-dimensional view is shown with materials noted. The geometrical and doping parameters are noted in the Methods simulation section.}
    \label{fig:Device}
\end{figure}

\begin{figure}[H]
    \includegraphics[width=\linewidth]{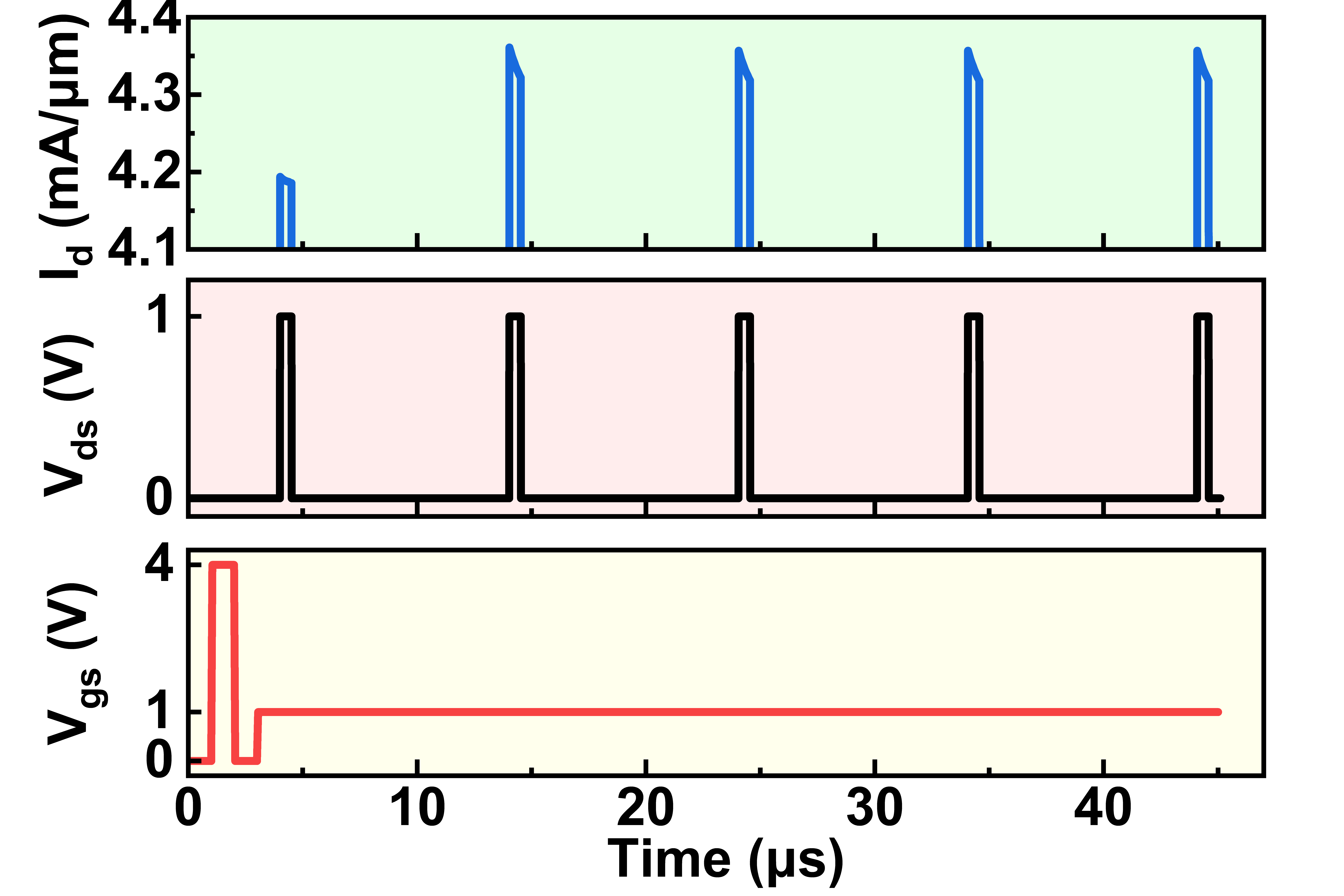}
    \caption*{\textbf{Supplementary Figure 8 \texttt{|} $I_d$ vs time simulation results of long IPI.}
    This simulation uses the same $V_{gs}$ scheme as Fig 6. The drain pulse inter-pulse interval (IPI) changes from 0.5 $\mu$s to 9.5 $\mu$s but retains its +1 V amplitude. The resulting $I_d$ exhibits no PPF after the first pulse.
    }
    \label{fig:LongIPI}
\end{figure}

\begin{figure}[H]
    \includegraphics[width=1\linewidth]{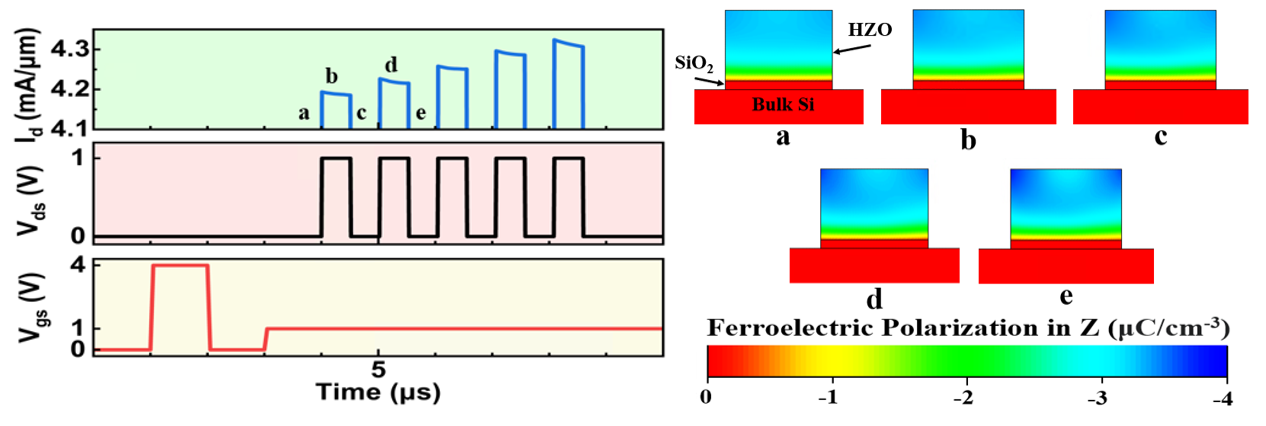}
    \caption*{\textbf{Supplementary Figure 9 \texttt{|} Evolution of ferroelectric polarization in the Z axis in gate polarization at +4 V.}
    The pulsing scheme shown was the input for the Ginestra$^{TM}$ simulations. An initial gate voltage pulse from 0 V to +4 V and back to 0 V precedes the final set at +1 V. 5 drain pulses are encoded between 0 V and +1 V with a duty cycle of 0.5 and a pulse width of 1 $\mu$s. The highlighted region of the $I_d$ subsection shows the locations (a-e) viewed using Ginestra$^{TM}$'s microscope tool to gain an insight into ferroelectric evolution. Ferroelectric polarization in the Z axis is mapped in a cross-sectional view of the gate region. a) Taken at 3.75 $\mu$s, after the gate setting operation but before the first drain pulse. b) Taken at 4.25 $\mu$s, at the top of the first drain pulse. c) Taken at 4.75 $\mu$s, between the first and second drain pulses. d) Taken at 5.25 $\mu$s, at the top of the second drain pulse. e) Taken at 5.75 $\mu$s, at 0 V after the second drain pulse. The evolution shows minimal change in polarization with time.
    }
    \label{fig:Ferro}
\end{figure}

\begin{figure}[H]
    \centering\includegraphics[width=0.75\linewidth]{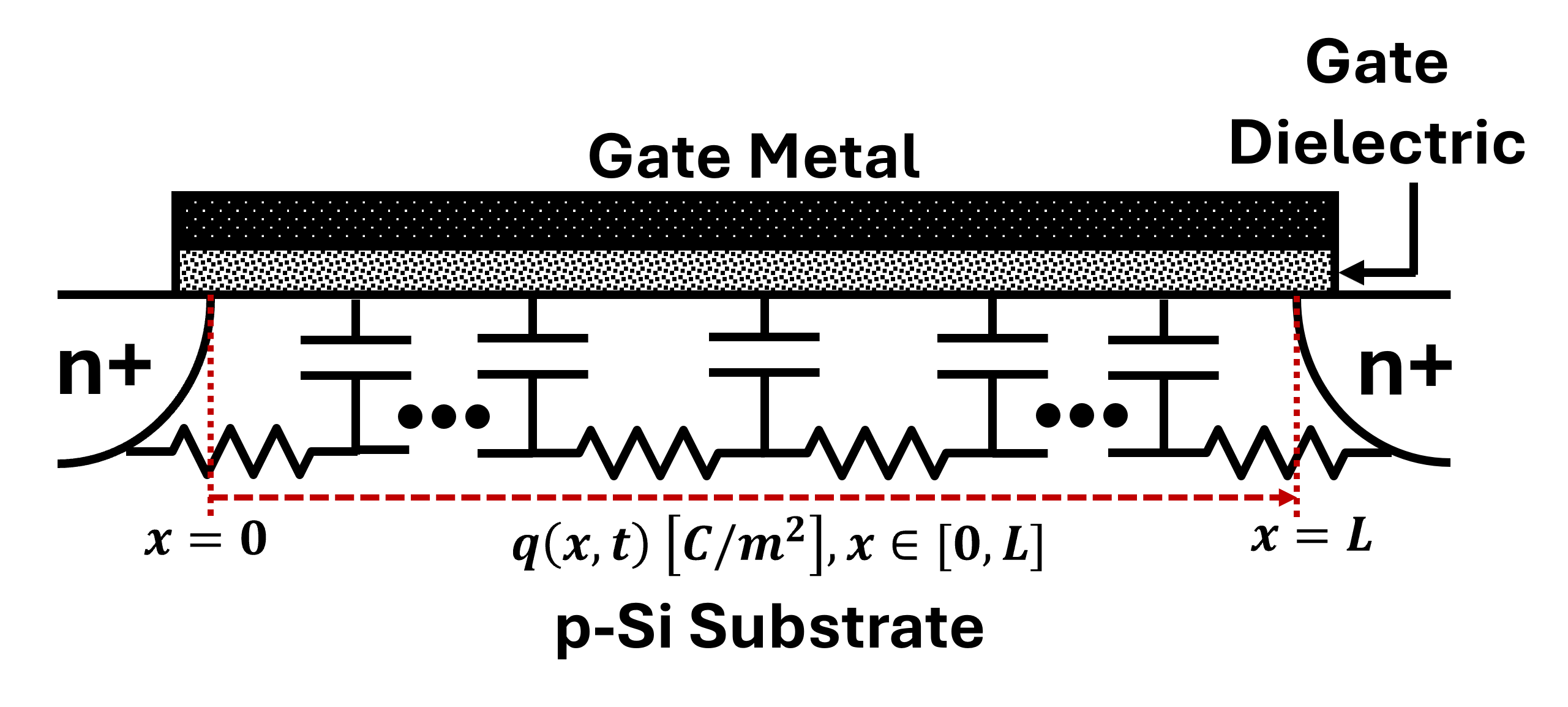}
    \caption*{\textbf{Supplementary Figure 10 \texttt{|} Equivalent RC network representing the Si MOSFET channel}
    Non-Quasi Static (NQS) model of the transistor is useful for explaining transient behavior, specially in rapid switching mode. The central idea for this model is the hypothesis that any transient voltage change at any terminal of a transistor do not result in an instantaneous change in the channel charge ($q(x,t)$) that gives rise to a transient component of the drain current. If the subsequent input pulse comes within the decay time of the transient current, the total current will change momentarily giving rise to short-term memory described in this paper.
    }
    \label{fig:LongIPI}
\end{figure}

\newpage
\section*{Supplementary Note: Mathematical formalism of the NQS effect}

\subsection*{Scope.}
This note summarizes a compact derivation of the non–quasi–static (NQS) response in MOSFETs under drain–voltage pulsing when a designed gate–source/drain overlap capacitance provides the input coupling. We follow the classical continuity–based transient treatment and its compact reduction to the “charge–deficit” model used in BSIM3/BSIM4, emphasizing quantities that appear in our experiments and simulations. \cite{OhWardDutton1980,PaulosAntoniadis1983,Cheng_NQS,BSIM4_manual}

\subsection*{S1. Charge–based baseline and continuity}
Let $q(x,t)$ [C/m$^2$] denote the inversion charge density along the channel ($x\!\in\![0,L]$), referenced to the operating point. Conservation of charge gives
\begin{equation}
\frac{\partial q(x,t)}{\partial t} + \frac{\partial i(x,t)}{\partial x} = 0,
\label{eq:continuity}
\end{equation}
where $i(x,t)$ is the lateral channel current density (A/m). Linearizing the constitutive relation about the bias point yields
\begin{equation}
i(x,t) \;\approx\; \sigma_{\! \mathrm{ch}} \frac{\partial \psi(x,t)}{\partial x}, 
\qquad
q(x,t) \;\approx\; C'_{\! \mathrm{ch}} \,\psi(x,t),
\label{eq:constitutive}
\end{equation}
with sheet conductivity $\sigma_{\! \mathrm{ch}}$ and channel charge susceptibility (capacitance per unit area) $C'_{\! \mathrm{ch}}$ that are \emph{bias dependent}. Eliminating $\psi$ gives a diffusion–type PDE for the incremental charge:
\begin{equation}
\frac{\partial q(x,t)}{\partial t} \;=\; D_{\! \mathrm{ch}} \,\frac{\partial^2 q(x,t)}{\partial x^2},
\qquad
D_{\! \mathrm{ch}} \;\equiv\; \frac{\sigma_{\! \mathrm{ch}}}{C'_{\! \mathrm{ch}}}.
\label{eq:diffusion}
\end{equation}
Equation~\eqref{eq:diffusion} is the continuum origin of NQS: the channel requires a finite time to redistribute charge after a fast perturbation. \cite{OhWardDutton1980}

\subsection*{S2. Boundary excitation via overlap capacitance}
A drain step $V_D(t) = V_{D0} + \Delta V_D\,u(t)$ couples into the gate/channel edge near $x=L$ through the gate–drain overlap capacitance $C_{gd,\mathrm{ov}}$. The injected displacement current is
\begin{equation}
i_{\mathrm{ov}}(t) = C_{gd,\mathrm{ov}}\,\frac{dV_D}{dt},
\label{eq:overlap_current}
\end{equation}
which imposes a boundary condition at $x=L$ equivalent to a localized charge injection/extraction that sets the initial condition $q(x,0^+)$ for Eq.~\eqref{eq:diffusion}. For compactness, we collect the distributed dynamics into an internal state $Q_{\mathrm{def}}(t)$ (the \emph{charge–deficit} variable).

\subsection*{S3. Modal solution and single–pole reduction}
Solving Eq.~\eqref{eq:diffusion} with standard mixed boundary conditions gives a modal expansion
\begin{equation}
q(x,t) = \sum_{n\ge1} A_n \,\phi_n(x)\, e^{-t/\tau_n},
\qquad
\tau_n \sim \frac{L^2}{(\pi n)^2 D_{\! \mathrm{ch}}}.
\label{eq:modes}
\end{equation}
The lowest pole $\tau_1$ dominates most device–level transients. Compact models therefore replace the distributed channel by a \emph{single–pole} surrogate with an internal state:
\begin{equation}
\frac{dQ_{\mathrm{def}}}{dt} = -\frac{Q_{\mathrm{def}}}{\tau} + S(t),
\qquad
I_k(t) = I_k^{\mathrm{DC}} + p_k\,\frac{dQ_{\mathrm{ch}}}{dt},
\label{eq:charge_deficit}
\end{equation}
where $S(t)$ encodes the excitation (e.g., Eq.~\eqref{eq:overlap_current}), $Q_{\mathrm{ch}}$ is the total channel charge, and $p_k$ are charge–conserving partition factors for terminals $k\in\{G,D,S\}$.\cite{OhWardDutton1980,PaulosAntoniadis1983,BSIM4_manual}

\subsection*{S4. The NQS time constant and its bias dependence}
In BSIM–style charge–deficit formulations, the dominant NQS time constant is written
\begin{equation}
\tau \;=\; R_{ii}\, W_{\mathrm{eff}}L_{\mathrm{eff}}\, C_{\mathrm{ox,eff}},
\label{eq:tau_BSIM}
\end{equation}
with $R_{ii}$ the \emph{intrinsic input resistance} that aggregates drift and diffusion contributions; $C_{\mathrm{ox,eff}}$ is the effective oxide capacitance per unit area.\cite{BSIM4_manual,Cheng_NQS} Because the diffusion component of $R_{ii}$ increases near threshold, 
\[
\tau(V_G)\; \text{is \emph{larger} in weak inversion (near/subthreshold) and \emph{smaller} in strong inversion},
\]
which is the established NQS bias dependence.\cite{BSIM4_manual,Cheng_NQS}

\subsection*{S5. Small–signal NQS (frequency domain)}
Linearizing Eq.~\eqref{eq:charge_deficit} gives first–order frequency responses; to leading order,
\begin{equation}
C_{gd}(\omega) \;\approx\; \frac{C_{gd,0}}{1 + j\omega\tau},
\qquad
g_m(\omega) \;\approx\; \frac{g_{m,0}}{1 + j\omega\tau_g},
\label{eq:freq_rolloff}
\end{equation}
capturing the roll–off of the effective drain–gate coupling and transadmittance with $j\omega\tau$. \cite{PaulosAntoniadis1983,BSIM4_manual} These forms are sufficient to fit the observed drain–step transients and AC corners.

\subsection*{S6. Drain–step transient and paired–pulse metrics}
For a single drain step, the terminal currents contain (i) a prompt displacement spike due to $C_{gd,\mathrm{ov}}$ and $C_{db}$, and (ii) a conduction relaxation with time constant $\tau$ set by Eq.~\eqref{eq:tau_BSIM}. For a \emph{paired–pulse} protocol with inter–pulse interval $T$ (IPI), the second pulse samples the residual channel–charge perturbation left by the first. A convenient scalar metric is the paired–pulse facilitation/depression (PPF/PPD) ratio, which—under broad conditions—follows
\begin{equation}
\mathrm{PPF/PPD}(T) \;\propto\; e^{-T/\tau},
\label{eq:PPF_law}
\end{equation}
where the sign depends on the operating point and excitation polarity (see S7). Equation~\eqref{eq:PPF_law} underpins our one–parameter fits of PPF vs. IPI.

\subsection*{S7. Operating–regime dependence (strong vs. weak inversion)}
Under a fixed external $V_G$, ferroelectric polarization selects the local regime at the drain edge. Positive polarization lowers the effective threshold $\Rightarrow$ \emph{strong inversion} (smaller $R_{ii}$, shorter $\tau$); negative polarization raises it $\Rightarrow$ \emph{weak inversion} (larger $R_{ii}$, longer $\tau$). \cite{BSIM4_manual,Cheng_NQS} In our biasing, the strong–inversion case exhibits paired–pulse \emph{facilitation}, while the weak–inversion case exhibits \emph{depression}. We interpret this polarity as consistent with NQS dynamics: the sign and magnitude of the residual channel–charge perturbation near the drain at time $T$ (just before the second pulse) depend on regime and partitioning; the shorter (longer) $\tau$ in strong (weak) inversion yields distinct residuals and hence PPF/PPD. \cite{OhWardDutton1980,PaulosAntoniadis1983} We emphasize that the polarity itself is \emph{operating–point specific} rather than universal.

\subsection*{S8. Practical extraction recipe (time domain)}
From a measured or simulated PPF–vs–IPI dataset $\{T_i,\mathrm{PPF}(T_i)\}$:
\begin{enumerate}[leftmargin=1.5em]
\item Subtract any baseline offset and fit $\mathrm{PPF}(T)$ to $A\,e^{-T/\tau}$ to obtain $\tau$ and its confidence interval.
\item Repeat for two polarization states (strong vs. weak inversion); $\tau$ should increase as the operating point moves toward weak inversion.
\item Optionally, vary $V_G$ in small steps around threshold to map $\tau(V_G)$ and corroborate the NQS bias trend.
\end{enumerate}

\subsection*{S9. Summary}
The NQS formalism follows directly from charge conservation and a distributed RC description of the inversion channel. Its compact reduction yields a single–pole time constant $\tau$ [Eq.~\eqref{eq:tau_BSIM}] that is \emph{predictably bias dependent}. In our overlap–engineered devices, drain–pulse experiments probe this $\tau$ via paired–pulse transients; the observed PPF/PPD and their timescales are consistent with the overlap–driven NQS mechanism. \cite{OhWardDutton1980,PaulosAntoniadis1983,Cheng_NQS,BSIM4_manual}




\end{appendices}
\newpage


\bibliography{sn-bibliography}% common bib file

\end{document}